\documentclass[twocolumn, preprintnumbers,amsmath,amssymb,amsfonts]{revtex4}
\usepackage{graphicx}
\usepackage{amsmath}
\usepackage{amssymb}
\usepackage{bm}
\usepackage{amsthm}
\usepackage{enumitem}

\begin{document}

	\title{Twenty-one characterizations of reversible quantum channels}

\author{Nan Li,$^{1,2,}$\footnote{linan@amss.ac.cn} Yuan Li,$^{3,}$\footnote{liyuan0401@snnu.edu.cn}  Hanyu Yang,$^{1,2,}$\footnote{yanghanyu@amss.ac.cn} and Shunlong Luo$^{1,2,}$\footnote{luosl@amt.ac.cn} }

\address{\vspace{0.1cm}$^1$State Key Laboratory of Mathematical Sciences, Academy of Mathematics and Systems Science, Chinese Academy of Sciences, Beijing 100190, China\\
$^2$School of Mathematical Sciences, University of Chinese Academy of Sciences, Beijing 100049, China\\
$^3$School  of  Mathematics  and
Statistics, Shaanxi Normal University,  Xi'an,
710062, China}

\begin{abstract}
Reversible quantum channels play a fundamental role in quantum dynamics and quantum information processing. A quantum channel is reversible if there exists another quantum channel acting as its left inverse.  Due to their intrinsic significance and wide applications, it is desirable to characterize reversible quantum channels from diverse perspectives.
In this work, we study reversible quantum channels on finite-dimensional Hilbert spaces, with particular emphasis on the case of different input and output dimensions. We systematically present twenty-one equivalent characterizations of reversible quantum channels from algebraic, geometrical, and information-theoretical perspectives. Among these characterizations, some are well known, while others, implicit in the literature or formulated in other contexts, are clarified here; the Choi-state characterization is derived in this work. Specifically, we prove that the Choi states of reversible quantum channels admit three equivalent forms: the spectral, direct-sum, and tensor-product representations. These twenty-one equivalent characterizations establish a comprehensive framework for reversible quantum channels, provide diverse insights into the structural and information-theoretic properties of quantum channels, and facilitate applications of reversibility in quantum information processing  such as quantum error correction, quantum teleportation, and quantum thermodynamics.
\end{abstract}

\maketitle

\section{Introduction}
 
 Reversibility and irreversibility are fundamental concepts in physics that describe whether a process can be perfectly undone.
 A reversible process can return to its initial state by a subsequent process without leaving any change in the system,  while an irreversible process proceeds in only one direction, resulting in   permanent changes  in various forms, such as state, information, or energy.

In quantum information theory, a quantum channel provides the most general description of a physically possible transformation that a quantum system can undergo \cite{Nielsen2012}. Mathematically, it is defined as a completely positive trace-preserving linear map  which transforms density operators from an input state space   to an output state space \cite{Choi1975,Kraus1983}.

A quantum channel $\mathcal{E}: \mathcal{S}({H}_a) \rightarrow \mathcal{S}({H}_b)$ is called reversible if there exists a quantum channel $\mathcal{R}: \mathcal{S}({H}_b) \rightarrow \mathcal{S}({H}_a)$ such that  $\mathcal{R}\circ \mathcal{E}(\rho)=\rho,\, \forall \, \rho \in \mathcal{S}({H}_a),$ where by convention, $\mathcal{S}({H})$  denotes the set of quantum states (density operators) acting on a Hilbert space ${H}$. 
Note that some literatures   focus on the reversibility of quantum channels with respect to two quantum states or a set of quantum states, which is  also referred to as the sufficiency or recoverability of quantum channels with respect to the given set.  Here, we use ``reversibility'' in the usual sense, meaning  that a quantum channel has a left inverse (which is itself a quantum channel) capable of reconstructing any state in the input state space.

The investigation of reversible quantum channels holds significant theoretical and practical value. Theoretically, it is connected to the data-processing inequality \cite{Lindblad1975,Uhlmann1977,Petz1986,Petz1988,Nielsen1997,Schumacher1996,Nielsen1998}, which is a cornerstone of quantum information theory. Practically, it has direct applications in  quantum teleportation \cite{Nielsen1997}, quantum error correction \cite{Schumacher1996,Nielsen1998,Caves1999,Knill1997,Knill2000},  quantum secret sharing \cite{Ogawa2005}, and quantum thermodynamics \cite{thermo0,thermo1, Characterizingirreversibility2018}.

Broadly speaking, characterizations of reversible (or  sufficient) quantum channels have been established in the literature  from the following two perspectives:

(1) From an information-theoretic (analytical) perspective, reversibility is closely linked to the preservation of certain  distinguishability measures.
These include quantities that can be expressed in terms of the quantum $f$-divergences \cite{Hiai2011,Hiai2017,Hiai2021},  such as the quantum relative entropy \cite{Petz1986,Petz1988,Ruskai2002,Petz2003,Petz2004,Hayden2004,Jencova2012,Shirokov2013}, the standard R\'{e}nyi divergences \cite{Petz1986quasi,Petz1986,Petz1988,JencovaPetz2006a,JencovaPetz2006b}, as well as quantities not expressible as the quantum $f$-divergences, such as  the sandwiched R\'{e}nyi divergences \cite{Jencova2017,Jencova2018,Jencova2021}, the quantum Fisher information \cite{Petz1996,JencovaPetz2006b}, the $\chi^2$-divergences \cite{Jencova2012}, the trace-norm distance \cite{Robin2008,Robin2010,Jencova2012,Jencova2024}, the quantum Chernoff distance \cite{Hiai2011,Hiai2017,Jencova2012,Audenaert2008},  and  the Hoeffding distance \cite{Hiai2011,Hiai2017,Jencova2012,Audenaert2008}.

(2) From a structural (algebraic) perspective,  reversible quantum channels  are described  based on the Knill-Laflamme theorem \cite{Knill1997,Knill2000}, the Petz recovery channel \cite{Petz1988,BarnumKnill2002}, the rotated Petz map \cite{Wilde2015,DupuisWilde2016,JungeWilde2016}, the Bayesian subjectivity  \cite{Liulizhuo2025}, the Choi state \cite{LuoSun2024,Liyuan2025}, the dual map \cite{Nielsen1998,LiYuan2025a,LiYuan2025b}, the complementary channel \cite{ Holevo2012,JFL2013}, the channel compatibility \cite{compatibility2017,GuoYi2025}, the representation theory \cite{Choi2009}, {\cite{Ticozzi2010}}, and preservation of operator orthogonality \cite{Busch1999,Molnar2002}.

In this work, we focus on reversible quantum channels on finite-dimensional Hilbert spaces, especially those with different dimensions for the input and output spaces, for the following reason.
Recall that the aim of quantum error correction is,  given   a quantum system described by a Hilbert space $H$ and a quantum noise operation described by a quantum channel $\mathcal{E}:\mathcal{S}(H)\rightarrow \mathcal{S}(H)$,  to identify a subspace $C$   (called the code) of the system space $ H$
 and a quantum operation capable of reversing the action of the noise on this code \cite{Knill1997,Knill2000}.
Therefore, the original quantum noise is usually modeled as a quantum channel whose  input and output state spaces have the same dimension. In fact, since we are only interested in the action of   $\mathcal{E}$ on $C$ and its inverse, we can restrict $\mathcal{E}$ to the code $C$ and only consider the  restricted  quantum channel  $\mathcal{E}|_C:\mathcal{S}(C)\rightarrow \mathcal{S}(H)$. Then, the code $C$ is correctable for $\mathcal{E}$ if and only if the restricted channel  $\mathcal{E}|_C$ is reversible \cite{Knill1997}.
Hence, it is essential to obtain characterizations of reversible quantum channels that have different dimensions for the input and output spaces in order to apply them effectively in quantum error correction theory.

The aim of this work is to present a transparent and unified framework for characterizing reversible quantum channels and to elucidate the interconnections among various characterizations. In particular, we systematically organize twenty-one equivalent characterizations, some of which are known in the literature while others are derived or clarified in this work. In particular, we establish a complete structural characterization of the Choi states for reversible quantum channels.

The remainder of this work is organized as follows. Section II introduces the notation and concepts used throughout the work. In Section III, we first present the definition of a reversible quantum channel in statement (R0), followed by a summary of twenty-one equivalent characterizations in statements (R1) through (R21). Among them, the Choi-state characterization (R7) is derived in this work. We also highlight the unique structure of recovery channels for reversible quantum channels. Section IV concludes with some discussions. 
Although many of these characterizations are known in the literature, some of the equivalences and their interconnections are not immediately apparent. For completeness and to provide a unified understanding of these results, we present detailed proofs of the equivalence between all these characterizations  in the appendix.

  \section{preliminaries}

We recall the notation, conventions, and basic definitions used throughout this work, including quantum channels, Choi states, dual maps, complementary channels, as well as the quantum relative entropy, quantum mutual information, and other related quantities.

Throughout this work, let  $\mathcal{L}(H_a,H_b)$ denote the Hilbert space of all linear operators from $H_a$ to $H_b$ (finite dimensional Hilbert space), equipped with the Hilbert-Schmidt inner product $\langle X|Y\rangle ={\rm tr}(X^\dagger Y).$
If $H_a=H_b=H$, we abbreviate $\mathcal{L}(H_a,H_b)$ as $\mathcal{L}(H)$.   Let  $\mathcal{S}(H)$ denote the  set of all  quantum states (density operators) on $H$.
For an operator $X$, let ${\rm rank}\,X$,  ${\rm spec}\,X$, and ${\rm supp}\,X$ denote its rank,  nonzero spectrum (nonzero eigenvalues with multiplicities), and support, respectively. Ler  $\Pi_X$ denote the projection onto ${\rm supp}\,X$.
The Schatten $p$-norm of $X$ is defined as
 \begin{align}
\|X\|_p\equiv ({\rm tr}|X|^p)^{\frac 1 p}, \quad p\geq 1,
\end{align}
where $|X|=\sqrt{X^\dagger X}$ is the absolute of $X$.

 Let
 \begin{align}\label{channelS}
{\cal E}: \mathcal{S}(H_a)\rightarrow \mathcal{S}(H_b)
 \end{align}
 denote a quantum channel (i.e., a completely positive trace-preserving  linear map), where $H_a$ and $H_b$ are the respective input and output Hilbert spaces with dimensions $d_{\alpha}={\rm dim}H_\alpha<\infty,$  and ${\bf 1}_\alpha$  denotes the identity operator on $H_\alpha$ for $\alpha=a,b$.
Due to linearity of ${\cal E}$, it can be naturally extended to a linear map
 \begin{align}\label{channelL}
{\cal E}: \mathcal{L}(H_a)\rightarrow \mathcal{L}(H_b).
 \end{align}
Throughout the work, we shall use the same symbol ${\cal E}$ for both maps, referring to Eq.  (\ref{channelS}) and Eq.  (\ref{channelL}) interchangeably.

Recall that any quantum channel ${\cal E}: \mathcal{S}(H_a)\rightarrow \mathcal{S}(H_b)$ has an operator-sum representation \cite{Choi1975,Kraus1983}
 \begin{align}
 \mathcal{E}(\rho )=\sum_j E_j\rho E_j^\dagger, \qquad \forall \ \rho \in \mathcal{S}(H_a),
 \end{align}
 with $E_j\in \mathcal{L}(H_a,H_b)$ such that $\sum_j E_j^\dagger E_j={\bf 1}_a$ (trace-preserving property).  Such a set $\{E_j\}$ is called a   Kraus representation of ${\cal E}.$
The Kraus representations of a quantum channel are not unique; there is a unitary freedom in the representation \cite{Nielsen2012}.
 If
   \begin{align}
   {\cal E}({\bf 1}_a)= \sum_j  E_jE_j^\dagger ={\bf 1}_b,
   \end{align}
  then ${\cal E}$  is called unital (identity-preserving).  Note that when $d_a\neq d_b$, there is no unital quantum channel from  $\mathcal{L}(H_a)$ to $\mathcal{L}(H_b)$ due to the trace-preserving property.
  Given a quantum channel ${\cal E}: \mathcal{S}(H_a)\rightarrow \mathcal{S}(H_b)$ as above, its dual map, or adjoint map, $\mathcal{E}^\dag: \mathcal{L}(H_b)\rightarrow \mathcal{L}(H_a)$ is defined as
   \begin{align}
  \mathcal{E}^\dag(Y )\equiv \sum_j E_j^\dag Y  E_j, \quad \forall \  Y\in \mathcal{L}(H_b).
   \end{align}
  It is obvious that for any quantum channel $\mathcal{E}$,  it holds that $\mathcal{E}^\dagger ({\bf 1}_b)={\bf 1}_a.$

   A quantum channel ${\cal E}: \mathcal{S}(H_a)\rightarrow \mathcal{S}(H_b)$ is called an isometric channel if  it can be written in the form $\mathcal{E}(\cdot)=V\cdot V^\dagger$, where $V\in \mathcal{L}(H_a, H_b)$ is an isometry, i.e., $V^\dagger V={\bf 1}_a.$

 According to the Stinespring dilation  theorem \cite{Stinespring1975}, any quantum channel $\mathcal{E}: \mathcal{S}(H_a)\rightarrow \mathcal{S}(H_b)$
can be represented as
 \begin{align}
 \mathcal{E}(\rho)={\rm tr}_{c}V\rho V^\dagger,\quad \forall\,  \rho\in\mathcal{S}(H_a),
 \end{align}
 where $V\in \mathcal{L}(H_a,H_b\otimes H_c)$ is an isometry. Then, the complementary channel
 \begin{align}
\widehat{\mathcal{E}}: \mathcal{S}(H_a)\rightarrow \mathcal{S}(H_c)
 \end{align}
 of ${\cal E}$ is defined as
 \begin{align}
\widehat{\mathcal{E}}(\rho )\equiv {\rm tr}_{b}V\rho V^\dagger,\quad \forall \, \rho\in\mathcal{S}(H_a).
 \end{align}
Note that the complementary channel is unique up to isometric equivalence, as described in Sec. 6.6 of \cite{Holevo2012}.

For  a quantum channel ${\cal E}: \mathcal{S}(H_a)\rightarrow \mathcal{S}(H_b)$,  its Choi state $J_{\cal E}\in  \mathcal{S}(H_a\otimes H_b)$ is defined as \cite{Choi1975}
 \begin{align}\label{Choi}
J_{\cal E}\equiv ( \mathcal{I}_a\otimes\mathcal{E})(|\Phi^+\rangle\langle \Phi^+|).
\end{align}
Here  $|\Phi^+\rangle=( \sum_{i=1}^{d_a}|i\rangle\otimes |i\rangle)/\sqrt{d_a}$ is a maximally entangled state on $H_a\otimes H_a$, and ${\cal I}_a: \mathcal{S}(H_a)\rightarrow \mathcal{S}(H_a)$ is the identity channel on  $H_a$. 
Throughout the work, we use the notation
$\{|i_\alpha\rangle:i=1,2,\cdots,d_\alpha\}$ to denote an orthonormal basis of $H_\alpha$. The subscript $\alpha$ specifies the Hilbert space $H_\alpha$ to which the basis belongs. For simplicity, we omit the subscript $a$ for the basis vectors of $H_a$ whenever no confusion can arise, while retaining the subscripts for all other Hilbert spaces. 

The Choi  operator   $C_{\mathcal{E}}$ of  ${\cal E}$   differs from the   Choi state $J_{\mathcal{E}}$ by a constant factor, i.e., 
 \begin{align}
C_{\cal E}\equiv \sum_{i,i'} |i\rangle \langle i'|\otimes {\cal E}(|i\rangle\langle i'|)  = d_a  J_{\cal E}.
 \end{align}

For two quantum channels $\mathcal{E}_k:\mathcal{S}(H_a)\rightarrow \mathcal{S}(H_{b_k}),  k=1,2,$
 their Jordan product is the map  \cite{Girard2021,GL2}
 \begin{align}
 \mathcal{E}_1\odot\mathcal{E}_2:\,\mathcal{L}(H_a)\rightarrow \mathcal{L}(H_{b_1}\otimes H_{b_2}), \label{JP1}
 \end{align}
 whose Choi  operator   is given by
\begin{align}
C_{\mathcal{E}_1\odot \mathcal{E}_2}\equiv \sum_{i,i',j,j'}\{|i\rangle\langle i'|, |j\rangle\langle j'|\}\otimes \mathcal{E}_1(|i\rangle\langle i'|)\otimes \mathcal{E}_2(|j\rangle\langle j'|), \label{JP2}
\end{align}
satisfying that ${\rm tr}_{b_1}C_{\mathcal{E}_1\odot \mathcal{E}_2}=C_{\mathcal{E}_2}$ and  ${\rm tr}_{b_2}C_{\mathcal{E}_1\odot \mathcal{E}_2}=C_{\mathcal{E}_1}$.
Here $\{|i\rangle\}$ is an orthonormal basis of $H_a$, and
$
\{X,Y\}= \frac 1 2 (XY+YX)
$ denotes the Jordan product of two operators $X$ and $Y.$ If the map $\mathcal{E}_1\odot\mathcal{E}_2$ is a quantum channel, i.e.,  $C_{\mathcal{E}_1\odot \mathcal{E}_2}\geq 0,$ then $\mathcal{E}_1$ and $\mathcal{E}_2$ are called Jordan-compatible.

The quantum data-processing inequality is a cornerstone of quantum information theory. A quantum state distinguishability measure $D(\rho\|\sigma)$ is said to satisfy the data-processing inequality if, for any quantum channel ${\cal E}: \mathcal{S}(H_a)\rightarrow \mathcal{S}(H_b)$, it obeys
\begin{align}\label{DPI}
D({\cal E}(\rho)\|{\cal E}(\sigma))\leq D(\rho\|\sigma),\qquad \forall \,\rho,\sigma\in \mathcal{S}(H_a).
\end{align}

In the following, we recall the definitions of several quantum state distinguishability measures, including  the quantum relative entropy, the standard R\'{e}nyi divergences,   the sandwiched R\'{e}nyi divergences,    the quantum Chernoff distance,  and  the Hoeffding distance. 

The quantum relative entropy between two states $\rho,\sigma\in \mathcal{S}(H)$  is defined as \cite{Umegaki1962,vedral2002relative}
\begin{align}
S(\rho\|\sigma)\equiv {\rm tr}\rho (\ln\rho-\ln\sigma).
\end{align}
Here  the logarithm is to the base $e$ (natural logarithm).

 For  $\alpha>0, \alpha \neq 1$, the standard R\'{e}nyi divergences between two states $\rho,\sigma\in \mathcal{S}(H)$  are defined as \cite{Petz1986quasi}
{\small
\begin{align}
D_{\alpha}(\rho\|\sigma)\equiv
\begin{cases}
\displaystyle
\frac{1}{\alpha-1}\ln\operatorname{tr}(\rho^{\alpha}\sigma^{1-\alpha}), & 0<\alpha<1,\\[1ex]
\displaystyle
\frac{1}{\alpha-1}\ln\operatorname{tr}(\rho^{\alpha}\sigma^{1-\alpha}), & \alpha>1,\ \operatorname{supp}\rho\subseteq\operatorname{supp}\sigma,\\[2ex]
+\infty, & \alpha>1,\ \operatorname{supp}\rho\nsubseteq\operatorname{supp}\sigma .
\end{cases}
\end{align}
}%
and the sandwiched R\'{e}nyi divergences are defined as \cite{Muller2013,Wilde2013}
{\small
\begin{align}
\widetilde{D}_{\alpha}(\rho\|\sigma) \equiv \bigg \{
\begin{array}{ll}
\displaystyle\frac{1}{\alpha-1}\ln\operatorname{tr}
 \big (\sigma^{\frac{1-\alpha}{2\alpha}}\rho
\sigma^{\frac{1-\alpha}{2\alpha}}\big )^{\alpha},  &
 \operatorname{supp} \rho \subseteq\operatorname{supp} \sigma \\
+\infty, & \text{otherwise.}
\end{array}
\bigg.
\end{align}
}%
Both of these two R\'{e}nyi divergences (R\'{e}nyi relative entropies) are defined for any $\alpha>0$ with $\alpha\neq 1$, while the values at $\alpha\in\{0,1,+\infty\}$ are derived via the corresponding limits. When $\alpha\rightarrow 1$, both of them yields the quantum relative entropy.

The Chernoff distance between two states $\rho,\sigma\in \mathcal{S}(H)$ is defined as \cite{Audenaert2008}
\begin{align}
C(\rho\|\sigma)\equiv\sup_{0\leq\alpha<1}(1-\alpha)D_{\alpha}(\rho \|\sigma),
\end{align}
and for any $r\in \mathbb{R}$, the Hoeffding distance is defined as \cite{Audenaert2008}
\begin{align}
H_r(\rho\|\sigma)\equiv\sup_{0\leq\alpha<1}\Big (\frac{-\alpha r}{1-\alpha}+D_{\alpha}(\rho \|\sigma)\Big ).
\end{align}

The Petz recovery map provides a complete characterization of the saturation of the quantum data-processing inequality.
Given a quantum channel ${\cal E}: \mathcal{S}(H_a)\rightarrow \mathcal{S}(H_b)$ and a quantum state $\gamma\in \mathcal{S}(H_a)$,  the action of the  Petz recovery channel $ \mathcal{R}^{\rm Petz}_{\mathcal{E},\gamma}: \mathcal{S}(H_b)\rightarrow \mathcal{S}(H_a)$ on ${\rm supp}{\cal E}(\gamma)$   is defined as  \cite{Petz1986,Petz1988,Petz2003,BarnumKnill2002}
\begin{align}\label{PetzChannel}
 \mathcal{R}^{\rm Petz}_{\mathcal{E},\gamma}(\sigma)   \equiv    \sqrt{\gamma}\ \mathcal{E}^\dagger\big (\sqrt{{\mathcal{E}(\gamma)}}^{-1} \sigma \,\sqrt{\mathcal{E}(\gamma)}^{-1}\big ) \sqrt{\gamma },
\end{align}
for any state  $\sigma\in {\cal S}(\Pi_{{\cal E}(\gamma)} H_b).$
We remark that whenever the inverse of an operator appears, it is understood to mean the Moore-Penrose pseudoinverse. For any $t\in \mathbb{R},$ 
the rotated Petz  map $ \mathcal{R}^{\rm Petz}_{{\cal E},\gamma;t}$ is defined as \cite{Wilde2015,DupuisWilde2016,JungeWilde2016} 
\begin{align}\label{rotatedPetz}
 \mathcal{R}^{\rm Petz}_{{\cal E},\gamma;t}(\sigma)   \equiv \big ( {\cal U}_{\gamma;-t}\circ  \mathcal{R}^{\rm Petz}_{\mathcal{E},\gamma}  \circ  {\cal U}_{{\cal E}(\gamma);t}\big )(\sigma), 
\end{align}
for any state  $\sigma\in {\cal S}(\Pi_{{\cal E}(\gamma)} H_b).$
Here  ${\cal U}_{\gamma;t}(\cdot )\equiv \gamma^{{\rm i}t }(\cdot) \gamma^{-{\rm i}t}$ is a partial isometric map, since $ \gamma^{{\rm i}t} \gamma^{-{\rm i}t}=\Pi_\gamma$, where ${\rm i}=\sqrt{-1}$ denotes the imaginary unit.
For notational simplicity,
the Petz recovery channel $ \mathcal{R}^{\rm Petz}_{\mathcal{E},{\bf 1}_a/{d_a}}$ is denoted by $
 \mathcal{R}^{\rm Petz}_{\mathcal{E}}$, and  
 the rotated Petz recovery map $ \mathcal{R}^{\rm Petz}_{\mathcal{E},{\bf 1}_a/{d_a};t}$ is denoted by $
 \mathcal{R}^{\rm Petz}_{\mathcal{E};t}$.  
 Acorrding to the  definitions,   $ \mathcal{R}^{\rm Petz}_{\mathcal{E}} $ and $
 \mathcal{R}^{\rm Petz}_{\mathcal{E};t}$ act  on any state  $\sigma\in {\cal S}(\Pi_{{\cal E}({\bf 1}_b)}H_b)$ as, respectively,
{\small
 \begin{align}
 \mathcal{R}^{\rm Petz}_{\mathcal{E}}  (\sigma)&=     \mathcal{E}^\dagger\big (\sqrt{{\mathcal{E}({\bf 1}_a)}}^{-1} \sigma \,\sqrt{\mathcal{E}({\bf 1}_a)}^{-1}\big ), \\
  \mathcal{R}^{\rm Petz}_{\mathcal{E};t}  (\sigma)&=     \mathcal{E}^\dagger\Big (\sqrt{{\mathcal{E}({\bf 1}_a)}}^{-1}{\cal E}\Big (\frac {{\bf 1}_a}{d_a}\Big)^{{\rm i}t} \sigma \,{\cal E}\Big(\frac {{\bf 1}_a}{d_a}\Big)^{-{\rm i}t}\sqrt{\mathcal{E}({\bf 1}_a)}^{-1}\Big ).
 \end{align}
 }
 
The quantum mutual information and the coherent information are important and widely used quantities in quantum information theory, both of which will be used in this work.
The quantum mutual information of a bipartite quantum state $\rho^{ab}\in {\cal S}(H_a\otimes H_b)$ is defined as \cite{vedral2002relative}
\begin{align}\label{mutualinformation}
I(\rho^{ab})\equiv S(\rho^{ab}\|\rho^{a}\otimes \rho^{b})=S(\rho^a)+S(\rho^b)-S(\rho^{ab}),
\end{align}
 where $\rho^a={\rm tr}_b\rho^{ab}$,  $\rho^b={\rm tr}_a\rho^{ab}$, and  $S(\rho)\equiv -{\rm tr}\rho \ln\rho$ denotes the von Neumann entropy of $\rho$.

For a quantum channel ${\cal E}: \mathcal{S}(H_a)\rightarrow \mathcal{S}(H_b)$ and a quantum state $\rho^a$ with purification $|\Psi_{a'a}\rangle$ on $H_{a'}\otimes H_a$,  the coherent information  is defined as \cite{Schumacher1996}
\begin{align}\label{coherentinformation}
I_c(\rho^{a},{\cal E})\equiv S\big({\cal E}(\rho^{a})\big)-S\big (({\cal I}_{a'}\otimes {\cal E})(|\Psi_{a'a}\rangle\langle  \Psi_{a'a}|)\big ),
\end{align}
which is just the negative conditional entropy of the bipartite state $({\cal I}_{a'}\otimes {\cal E})(|\Psi_{a'a}\rangle\langle  \Psi_{a'a}|).$ Here ${\cal I}_{a'}$ denotes the identity channel on $H_{a'}.$

\vskip  0.2cm
\section{Characterizations of reversible quantum channels}

In this section, we provide a systematic study of reversible quantum channels through twenty-one equivalent characterizations, as stated in the following proposition. We take (R0) as the definition and prove that all subsequent statements (R1)-(R21) are equivalent to (R0).
Among these characterizations, some are  well known, such as the Knill-Laflamme condition (R1), the Petz-recovery-map characterization (R4), and the preservation of pairwise distinguishability measure (R18), which is closely related to the data-processing inequality. 
Some are either implicit in existing results or originally formulated in different contexts rather than specifically in terms of reversible quantum channels, such as (R12)-(R17). We provide the corresponding references for  these  characterizations; however, the direct proofs of these equivalences may not be readily available in the references. For completeness, we provide    detailed proofs of all equivalences in the appendix.
In addition to organizing these known and reformulated characterizations into a unified framework, we derive the Choi-state characterization (R7) of reversible quantum channels and present its three equivalent representations.

The characterizations are organized as follows. Statements (R1)-(R16) provide structural (algebraic) characterizations of reversible quantum channels. In particular, (R2), (R3), and (R8) are closely related to the Knill-Laflamme condition (R1), while (R4)-(R6) are associated with the Petz recovery map. Statements (R7)-(R9) are formulated in terms of the Choi states, either of reversible quantum channels or of their complementary channels. Moreover, (R8)-(R11) and (R20) are connected to complementary channels.
The remaining algebraic characterizations (R12)-(R16) describe reversible channels from various operator-theoretic perspectives. Statement (R17) gives a geometric characterization based on Schatten $p$-norm preservation. Finally, (R18)-(R21) provide information-theoretic (analytical) characterizations in terms of the preservation of suitable information measures.

\vskip 0.2cm

\noindent{\bf Proposition.} For a quantum channel $\mathcal{E}: \mathcal{S}(H_a)\rightarrow \mathcal{S}(H_b)$, the following statements (R0)-(R21) are equivalent.

\begin{enumerate}[label=(R\arabic*), start=0]
\item \label{R0}
(\textit{Existence of left-inverse channel}) There exists  a quantum channel $\mathcal{R}:\mathcal{S}(H_b)\rightarrow \mathcal{S}(H_a)$ such that
\begin{align}
\mathcal{R}\circ \mathcal{E}=\mathcal{I}_a.
\end{align}
In this case, it must  hold that $d_a\leq d_b$, and  $\mathcal{R}$ is called a recovery channel of ${\cal E}.$
\vskip 0.2cm
\item  \label{RKL}
(\textit{Knill-Laflamme condition})  For any Kraus representation $\{E_j:j=1,2,\cdots,n\}$ of ${\cal E},$ there exist  $\lambda _{jj'}\in \mathbb{C}$ for $j,j'=1,2,\cdots,n$ such that  \cite{Knill1997}
\begin{align}
E_j^\dagger E_{j'}=\lambda_{jj'}\,{\bf 1}_a.
\end{align}
In this case, taking trace on both sides gives   $ \lambda_{jj'}={\rm tr}(E_j^\dagger E_{j'})/d_a.$
\vskip 0.2cm
\item \label{Rrandomiso}
 (\textit{Orthogonal isometric decomposition})
 There exist a probability distribution $\{\lambda_k:k=1,2,\cdots,m\}$ with $\lambda_k>0$ and  $\sum_{k=1}^m \lambda_k = 1$, and a set of orthogonal isometries $\{V_k\in \mathcal{L}(H_a,H_b):k=1,2,\cdots,m\}$ (i.e., $V_k^\dagger V_{k'}=\delta_{kk'}\,{\bf 1}_a$) such that   \cite{Busch1999,LiYuan2025b}
\begin{align}\label{OrtIso}
\mathcal{E}(\rho)=\sum_{k=1}^m \lambda_k\, V_k\rho V_k^\dagger,
\end{align}
for any state $ \rho\in \mathcal{S}(H_a).$
In this case, $m={\rm rank}\, J_{\mathcal{E}}, $ and $ \{\lambda_1,\lambda_2,\cdots,\lambda_m\}={\rm spec}\, J_{\mathcal{E}}.$ This characterization  implies that $m \leq \lfloor d_b/d_a\rfloor$, where $\lfloor x \rfloor$ denotes the greatest integer upper bounded by  $x$.
\vskip 0.2cm
\item \label{Rdual}
 (\textit{Scaled dual-map inversion}) The dual map $\mathcal{E}^\dagger$ satisfies \cite{Nielsen1998,LiYuan2025a,LiYuan2025b}
\begin{align}
\mathcal{E}^\dagger \circ \mathcal{E}=\beta\,{\cal I}_a
\end{align}
for some $\beta \in (0,1].$  In this case, direct calculation shows that $\beta={\rm tr}J_{\mathcal{E}}^2={\rm tr}\big (\mathcal{E} ({\bf 1}_a)^2\big )/d_a$.
\vskip 0.2cm
\item  \label{RPetz}
(\textit{Petz-recovery-map characterization}) The Petz recovery map $ \mathcal{R}^{\rm Petz}_{\mathcal{E}} $ satisfies  \cite{Petz2004,BarnumKnill2002}
\begin{align}
\mathcal{R}^{\rm Petz}_{\mathcal{E}}  \circ \mathcal{E}= {\cal I}_a.
\end{align}
\vskip 0.2cm
\item \label{RrotatedPetz}
(\textit{Rotated-Petz-map characterization}) The  rotated Petz map $  \mathcal{R}^{\rm Petz}_{{\cal E};t} $ satisfies \cite{Wilde2015} 
\begin{align}
 \mathcal{R}^{\rm Petz}_{{\cal E};t}  \circ \mathcal{E}= {\cal I}_a,\quad \quad \forall\, t\in  \mathbb{R}.
\end{align}
\item \label{RBayesian}
(\textit{Prior-state independence of Petz recovery maps})
Given two arbitrary prior states $\gamma_1,\gamma_2\in {\cal S}(H_a),$ the corresponding Petz recovery maps $\mathcal{R}^{\rm Petz}_{\mathcal E,\gamma_{1}}$  and  $\mathcal{R}^{\rm Petz}_{\mathcal E,\gamma_{2}}$ satisfy that
\begin{align}
\mathcal{R}^{\rm Petz}_{\mathcal E,\gamma_1}\circ {\cal E}(\rho)
=
\mathcal{R}^{\rm Petz}_{\mathcal E,\gamma_2}\circ {\cal E}(\rho),
\end{align}
 for any state $\rho\in {\cal S}(H_a)$  such that ${\rm supp}\rho \subseteq {\rm supp}\gamma_1\cap {\rm supp}\gamma_2$   \cite{Petz2004,Liulizhuo2025}.
\vskip 0.2cm
\item \label{RChoi}
(\textit{Choi-state characterization}) 
The Choi state  $J_{\mathcal{E}}$ admits the following three equivalent characterizations. First, it admits the  following  spectral decomposition:
\begin{align}\label{Choisum}
J_{\mathcal{E}} = \sum_{k=1}^m \lambda_k \, |\Psi_k\rangle\langle\Psi_k|.
\end{align}
Here  $\lambda_k> 0$ with $\sum_{k=1}^m \lambda_k = 1$, and eigenvectors $|\Psi_k\rangle\in {H}_a \otimes {H}_b$ satisfying
$
\operatorname{tr}_b(|\Psi_k\rangle\langle\Psi_{k'}|) =\delta_{kk'} { \mathbf{1}_a}/{d_a}.
$ 
Second,  the above spectral decomposition is equivalently written as the following direct-sum decomposition: 
\begin{align}\label{Choidirectsum}
J_\mathcal{E}
=
\bigoplus_{k=1}^m
\lambda_k
|\Phi_k^+\rangle\langle\Phi_k^+|.
\end{align}
Here 
$
|\Phi_k^+\rangle
=(\sum_{i=1}^{d_a}
|i\rangle\otimes|i_{b_k}\rangle)/{\sqrt{d_a}}
$
is a  maximally entangled state on $H_a\otimes H_{b_k}$. $H_b$ has the decomposition
$
H_b=
 (\oplus_{k=1}^mH_{b_k} )\oplus H_{b_0}$ with $H_{b_k}\cong H_a.$ 
 Finally, under the natural identification
$\bigoplus_{k=1}^{m} H_{b_k} \cong H_a \otimes H_c$,
where $\dim H_c=m$, the above direct-sum decomposition  can be equivalently expressed in the following tensor-product representation:  
 \begin{align}\label{Choitensor}
J_\mathcal{E}
=({\bf 1}_a\otimes U_b) ( |\Phi^+\rangle\langle\Phi^+|\otimes \tau) ( {\bf 1}_a\otimes U_b^\dagger).
\end{align}
Here $\tau\in {\cal S}(H_c)$,  $U_b$ is a unitary operator on $H_b$, 
and  $|\Phi^+\rangle=( \sum_{i=1}^{d_a}|i\rangle\otimes |i\rangle)/\sqrt{d_a}$ is a maximally entangled state on $H_a\otimes H_a$.
For simplicity, we do not distinguish between an operator on a direct-sum subspace and its zero extension to the whole Hilbert space. Accordingly, the zero operator on the remaining subspace may be omitted whenever no confusion arises.

\vskip 0.2cm
\item  \label{RKLoperator}
(\textit{Block-matrix-form Knill-Laflamme condition})  Any Kraus representation $\{E_j:j=1,2,\cdots,n\}$ of ${\cal E}$ satisfies 
\begin{align}
\sum_{j,j'=1}^n|j_c\rangle\langle j'_c|\otimes E_j^\dagger E_{j'}=\Lambda\otimes {\bf 1}_a.
\end{align}
Here $\Lambda=\sum_{jj'} \lambda_{jj'}|j_c\rangle \langle j'_c| \in {\cal L}(H_c)$ 
 with $\lambda_{jj'}={\rm tr}(E_j^\dagger E_{j'})/{d_a}$ and $\{|j_c\rangle:j=1,2,\cdots,n\}$ being an orthonormal basis of $H_c=\mathbb{C}^n.$
\vskip 0.2cm
\item \label{RChoiCom}
(\textit{Product-form complementary Choi state}) The Choi state  $J_{\widehat{\mathcal{E}}}$ of the complementary channel  $ {\widehat{\mathcal{E}}}$ is a product state, i.e., \begin{align}\label{R9}
J_{\widehat{\mathcal{E}}}=\frac{{\bf 1}_a}{d_a}\otimes \widehat{\mathcal{E}}\Big (\frac{{\bf 1}_a}{d_a}\Big ). 
\end{align}
\vskip 0.2cm
\item   \label{Rreplacement}
(\textit{Complementary to replacement channel})
The complementary channel $\widehat{\mathcal{E}}:\mathcal{L}(H_a)\rightarrow \mathcal{L}(H_c)$ is a replacement channel, i.e., there exists a state $\tau\in \mathcal{S}(H_c)$ such that 
\begin{align}
\widehat{\mathcal{E}}(X)=({\rm tr}X)\, \tau,\quad\quad \forall\, X\in \mathcal{L}(H_a).
\end{align}
\vskip 0.2cm
\item  \label{RJordan} 
(\textit{Jordan-compatibility between the  identity channel and the complementary  channel})
$\mathcal{I}_a$ and  $\widehat{\mathcal{E}}$ are Jordan-compatible, i.e.,  \cite{compatibility2017,GuoYi2025}
\begin{align}
C_{\mathcal{I}_a\odot\widehat{\mathcal{E}}}\geq 0.
\end{align}
Here,  $\mathcal{I}_a\odot\widehat{\mathcal{E}}$ is the Jordan product of  $\mathcal{I}_a$ and  $\widehat{\mathcal{E}},$ as defined by Eqs. (\ref{JP1}) and (\ref{JP2}).
\vskip 0.2cm
\item   \label{Rappendingunitary} (\textit{Isometric-appending decomposition})
There exist a decomposition
$
H_b=(H_a\otimes H_{c})\oplus H_{b_0},
$ a diagonal state
$\tau=\operatorname{diag}(\lambda_1,\cdots,\lambda_m)
\in\mathcal S(H_{c}),
$ and an isometry $W:H_a\otimes H_{c}\rightarrow H_b$ such that \cite{Busch1999,LiYuan2026}
\begin{align}\label{Isodecomp}
\mathcal E(\rho)
&=W(\rho\otimes\tau)W^\dagger .
\end{align}
Equivalently, by extending the isometry $W$ to a unitary operator
$U$ on $H_b$, the channel $\mathcal{E}$ can be decomposed as
\begin{align}\label{Unidecomp}
\mathcal{E}
=
\mathcal{U}\circ\mathcal{T}_{\tau},
\end{align}
where
$
\mathcal{T}_{\tau}(\rho)
=
(\rho\otimes\tau)\oplus{\bf 0}_{b_0},
  \forall\,\rho\in\mathcal{S}(H_a),
$
is an appending channel, and
$\mathcal{U}(\sigma)
=
U\sigma U^\dagger,
  \forall\,\sigma\in\mathcal{S}(H_b),
$ is a unitary channel.
\vskip 0.2cm
\item  \label{Rrepresentation}
(\textit{Representation up to smearing by a fixed operator})
 There is a representation $\pi:\mathcal{L}(H_a)\rightarrow \mathcal{L}(H_b)$ (i.e., a linear
multiplicative map preserving the adjoint operation) such that \cite{Choi2009}
\begin{align}
 \mathcal{E}(X)=\pi (X)\mathcal{E}({\bf 1}_a)=\mathcal{E}({\bf 1}_a) \pi (X),
\end{align}
for any operator $X\in \mathcal{L}(H_a).$
\vskip 0.2cm
\item  \label{Roperatormultiply}
(\textit{Multiplication preservation up to an operator})  For any two  states  $\rho,\sigma\in \mathcal{S}(H_a)$,  it holds that
\begin{align}
 \mathcal{E}(\rho)\mathcal{E}(\sigma)=\mathcal{E}(\rho\sigma)\mathcal{E}({\bf 1}_a)=\mathcal{E}({\bf 1}_a)\mathcal{E}(\rho\sigma).
\end{align}
\vskip 0.2cm
\item \label{Rvectorortho}
 (\textit{Preservation of pure-state orthogonality})   For any two orthogonal pure states $|\psi\rangle,|\varphi\rangle \in  H_a$ (i.e.,  $\langle \psi |\varphi \rangle=0$),  it holds that \cite{Busch1999}
\begin{align}
\mathcal{E}(|\psi\rangle\langle \psi|)\mathcal{E}(|\varphi\rangle\langle \varphi|)=0.
\end{align}
\vskip 0.2cm
\item  \label{Roperatorortho}
(\textit{Preservation of operator orthogonality})   For any two states $\rho,\sigma\in \mathcal{S}(H_a)$ such that $\rho\sigma=0$,  it holds that \cite{Busch1999}
\begin{align}
\mathcal{E}(\rho)\mathcal{E}(\sigma)=0.
\end{align}
\vskip 0.2cm
\item  \label{RSchatten}
(\textit{Preservation of Schatten $p$-norm})
There exist some $p\geq1$ and a corresponding constant 
$\gamma_p\in(0,1]$ such that \cite{LiYuan2025a}
\begin{align}
\|\mathcal{E}(X)\|_p
=
\gamma_p\|X\|_p, 
\end{align}
for any operator $X\in \mathcal{L}(H_a).$
For this value of $p$, the constant $\gamma_p$ is given by
$\gamma_p=\|J_{\mathcal{E}} \|_p=\big(\sum_{k=1}^m \lambda_k^p\big)^{1/p}.$
\vskip 0.2cm
\item  \label{RDPI}
(\textit{Preservation of pairwise distinguishability measure})  It holds that
\begin{align}
D\big (\mathcal{E} (\rho) \|\mathcal{E}  (\sigma) )=D\big(\rho  \|\sigma),
\end{align}
for any states $\rho,\sigma  \in \mathcal{S}(H_a).$
 Here $D(\rho\|\sigma)$ denotes any state distinguishability measure satisfying the
data-processing inequality and such that equality in the inequality implies
the existence of a recovery map.
Such measures include the quantum relative entropy $S(\rho\|\sigma)$  \cite{Petz1986,Petz1988},  the standard R\'{e}nyi relative entropy $D_\alpha(\rho\|\sigma)$ for
 some  $\alpha\in (0,1)\cup(1,2)$ \cite{JencovaPetz2006a,JencovaPetz2006b},  the sandwiched R\'{e}nyi relative entropy  $\widetilde{D}_\alpha(\rho\|\sigma)$ for some $\alpha\in (1/2,1)\cup(1,+\infty)$ \cite{Jencova2017,Jencova2018,Jencova2021},
  the Chernoff distance $C(\rho\|\sigma)$, or  the Hoeffding distance $H_r(\rho\|\sigma)$ for all $r\in \mathbb{R}$ \cite{Jencova2012}.
\vskip 0.2cm
\item \label{Rtotalcorrelations}  
(\textit{Preservation of  correlations in maximally entangled  states})
It holds that \cite{JFL2013}
\begin{align}
I(J_{\cal E})=I(|\Phi^+\rangle\langle \Phi^+|).
\end{align}
Here $I(\rho)$ denotes the quantum mutual information of a bipartite quantum state $\rho$ defined by Eq. (\ref{mutualinformation}).
 \vskip 0.2cm
\item \label{Rcomplementary} 
(\textit{No correlations are transferred to the complementary channel}) The Choi state of the complementary channel has vanishing total correlations, namely, \cite{luo2011decoherence}
\begin{align}
I(J_{\widehat{{\cal E}}})=0.
\end{align}
\item \label{Rcoherent} 
(\textit{Coherent-information characterization})
It holds that \cite{Schumacher1996}
\begin{align}
S\Big (\frac {{\bf 1}_a}{d_a}\Big )=I_c\Big (\frac {{\bf 1}_a}{d_a}, {\cal E}\Big ).
\end{align}
Here $I_c( {{\bf 1}_a}/{d_a},{\cal E})$ is the quantum coherent information with respect to ${\cal E}$ and ${{\bf 1}_a}/{d_a}$ defined by Eq. (\ref{coherentinformation}).
\end{enumerate}

 \vskip 0.3cm
Finnaly, we emphasize the unique structure of recovery channels for reversible quantum channels, as established in \cite{LiYuan2025b}. Specifically,
for a reversible quantum channel ${\cal E}: \mathcal{S}(H_a)\rightarrow \mathcal{S}(H_b)$,
 a quantum channel  $ \mathcal{R}: \mathcal{S}(H_b)\rightarrow \mathcal{S}(H_a)$ is a recovery channel if and only if it has the form \cite{LiYuan2025b}
\begin{align}
\mathcal{R}(\sigma)& =\mathcal{R}^{\rm Petz}_{\mathcal{E}}(P\sigma P)+\mathcal{R}_{\perp}\big (P_{\perp}\sigma P_{\perp}\big ),
\end{align}
for any state $\sigma\in {\cal S}(H_b).$
Here $P=\Pi_{{\cal E}({\bf 1}_a) }$, $P_{\perp}={\bf 1}_b-P$, and  
$
\mathcal{R}_{\perp}:\mathcal{S}(P_{\perp}H_b)\rightarrow \mathcal{S}(H_a)
$
is an arbitrary channel.
For    $\mathcal{E}(\rho)=\sum_{k=1}^m \lambda_k\, V_k\rho V_k^\dagger$ as   in   (R2), the Petz recovery channel acts on any state  $\sigma   \in \mathcal{S}(PH_b)$ as
\begin{align}
\mathcal{R}^{\rm Petz}_{\mathcal{E}}(\sigma) =\mathcal{E}^\dagger  \big (  {\sqrt{ \mathcal{E}({\bf 1}_a)}}^{-1}\sigma {\sqrt{ \mathcal{E}({\bf 1}_a )}}^{-1} \big )=&\sum_{k=1}^m  V_k^\dagger \sigma  V_k.
\end{align}

\section{Discussion}

A quantum channel is reversible only if the input dimension does not exceed the output dimension. It is well known that when the input and output dimensions are equal, every reversible quantum channel is unitary. In contrast, when the input dimension is strictly smaller than the output dimension, a reversible quantum channel is generally a convex combination of orthogonal isometric channels. This more general setting is particularly relevant to quantum error correction. Motivated by this distinction, we have focused  on reversible quantum channels with possibly different input and output dimensions and have  systematically established the equivalence of twenty-one characterizations of such channels. These include well-known characterizations, reformulations of implicit existing results within a unified framework, and new characterizations established in this work.
These characterizations span multiple mathematical perspectives, revealing the rich structure underlying quantum reversibility.

From an algebraic-structural perspective, reversible quantum channels admit a canonical decomposition as convex combinations of orthogonal isometric channels (R2), which follows directly from the Knill-Laflamme condition formulated in terms of Kraus operators (R1). This fundamental structure gives rise to several equivalent algebraic descriptions, including the scaled dual-map inversion condition (R3) and the block-matrix formulation of the Knill-Laflamme condition (R8). The Petz recovery map plays a central role in determining the structure of recovery channels, leading to the Petz-recovery-map characterization (R4), the rotated-Petz-map characterization (R5), and the prior-state independence of Petz recovery maps (R6).

Motivated by the channel-state duality, reversible quantum channels can also be characterized through their Choi states. In particular, we have  obtained a complete structural characterization of Choi states for reversible quantum channels in (R7). The complementary-channel perspective provides another viewpoint on reversibility and leads to several equivalent characterizations, including the product-form structure of complementary Choi states (R9), the replacement-channel structure of complementary channels (R10), and the Jordan compatibility between the identity channel and the complementary channel (R11). Other algebraic characterizations include the decomposition into  isometric and appending channels (R12), the representation up to smearing by a fixed operator (R13), the preservation of operator multiplication up to an operator (R14), and the preservation of orthogonality for pure states (R15) and general operators (R16). These structural characterizations provide deep insights into the internal architecture of reversible quantum channels and their intimate connection with isometric embeddings.

From an information-theoretic perspective, reversible quantum channels are precisely those that preserve Schatten $p$-norms (R17) and a broad class of distinguishability measures (R18), including the quantum relative entropy, standard and sandwiched Rényi divergences, the Chernoff distance, and the Hoeffding distance. The preservation of Schatten $p$-norms also provides a geometric characterization of reversibility, and is closely related to various capacities of quantum channels \cite{Kuperberg2003}, highlighting its information-theoretic significance. Furthermore, reversible channels preserve correlations in an information-conserving manner: the total correlations of the initial maximally entangled state are entirely transferred to the Choi state of the channel (R19) and remain absent from the Choi state of the complementary channel (R20). Moreover, reversible channels preserve the coherent information (R21), which quantifies the quantum information transmitted through a channel and plays a fundamental role in quantum error correction and quantum communication.
These information-theoretic characterizations establish a direct connection between reversibility and fundamental principles in quantum information theory, including the saturation of the data-processing inequality, the information conservation principle, and the complementarity principle.
 
The interconnections between these characterizations, as demonstrated through the detailed proofs in the appendix, reveal the deep unity underlying apparently disparate mathematical descriptions. The structure of recovery channels on the support of
${\cal E}({\bf 1}_a)$ is uniquely determined by the Petz recovery channel, while its action on the orthogonal complement remains arbitrary, a flexibility that has implications for approximate quantum error correction. 
 
This multifaceted review not only synthesizes results scattered across the literature but also provides a unified framework for understanding reversible quantum processes. The twenty-one characterizations offer complementary tools for investigating reversibility in various contexts: algebraic conditions for structural analysis, analytical conditions for information-theoretic applications, and geometrical conditions for intuitive visualization. We anticipate that these characterizations will find applications in quantum error correction, quantum thermodynamics, and the study of quantum dynamics. Future work may extend these results to infinite-dimensional systems and explore approximate versions of reversibility relevant for practical quantum information processing. In particular, the issue of quantification of irreversibility is of fundamental importance and practical usage, and is worth further investigation from as many angles as possible.

\section*{Acknowledgment}

This work was supported by Beijing Natural Science Foundation, Grant No. Z250004 and the National Natural Science Foundation of China, Grant No. 12341103.


\vskip 0.5cm
\begin{center} 
{\bf APPENDIX}
\end{center}
Here we present detailed proofs of the proposition establishing the equivalence of (R0)-(R20).
 \vskip 0.15cm

 First, we prove that {\bf (R0)$\Rightarrow$(R1)$\Rightarrow$(R2)$\Rightarrow$(R0)}.
 \vskip 0.15cm

{\bf (R0)$\Rightarrow$(R1)}

Suppose that $\mathcal{E}:\mathcal{S}(H_a)\rightarrow\mathcal{S}(H_b)$ is reversible and $\mathcal{R}:\mathcal{S}(H_b)\rightarrow\mathcal{S}(H_a)$ is a recovery  channel of $\mathcal{E}$, then
\begin{align}
\mathcal{R}\circ\mathcal{E}=\mathcal{I}_a.
\end{align}
Let $\{E_j\}$ and  $\{R_i\}$ be  Kraus representations  of $\mathcal{E}$ and $\mathcal{R}$, respectively. Then  $\{R_iE_j\}$ forms  a Kraus representation  of the identity channel $\mathcal{I}_a$. 
The Kraus operators of the identity channel span a one-dimensional space. Hence,
\begin{align}
R_iE_j\in {\rm span}\{{\bf 1}_a\},\quad {\rm i. e.,}\quad   R_iE_j=\alpha_{ij}{\bf 1}_a,
\end{align}
with $\alpha_{ij}={\rm tr}(R_iE_j)/d_a\in \mathbb{C}.$ Using $\sum_i R_i^\dagger R_i={\bf 1}_b,$  we get  that
\begin{align}\nonumber
E_j^\dagger E_{j'}&=E_j^\dagger \Big  (\sum_i R_i^\dagger R_i\Big ) E_{j'}\\ \nonumber
&=\sum_i  ( R_i E_{j})^\dagger   ( R_i E_{j'})\\
&=\Big (\sum_i   {\alpha^*_{ij}}\alpha_{ij'}\Big ){\bf 1}_a.
\end{align}
Here $x^*$ denotes the conjugate of a complex number $x$.
Define
$\lambda_{jj'}= \big (\sum_i   {\alpha^*_{ij}}\alpha_{ij'}\big )$.
This completes the proof of (R0)$\Rightarrow$(R1).
\vskip 0.2cm

{\bf (R1)$\Rightarrow$(R2)}

  Suppose that (R1) holds. Following the notations in the above deduction, it is easy to verify that the $n\times n$ complex Gram matrix $\Lambda= (\lambda_{jj'})$ is a trace-one positive semidefinite matrix. Thus, $\Lambda$ can be diagonalized as $D=U^\dagger \Lambda U,$ where $U=(u_{ij})$ is a unitary matrix, and   $D={\rm diag} (\lambda_1,\lambda_2,\cdots ,\lambda_m,0,\cdots,0)$ is a diagonal matrix with $\{\lambda_1,\lambda_2,\cdots,\lambda_m\}={\rm spec}\,\Lambda$ being the nonzero spectrum of $\Lambda$ counted with multiplicity.   Let
\begin{align}
F_k\equiv\sum_{j=1}^n  {u_{jk}}E_j, \qquad  k=1,2,\cdots,m,
\end{align}
with $m={\rm rank}\,\Lambda={\rm rank}\,J_{\cal E}$ the Choi rank of ${\cal E}$.
Then it can be directly verified that  $\{F_k: k=1,2,\cdots,m\}$ is another Kraus representation of $\mathcal{E}$  satisfying that
\begin{align}
F_k^\dagger F_{k'}=\delta_{kk'}\lambda_k\, {\bf 1}_a, \qquad  k,k'=1,2,\cdots,m.  \label{C1}
\end{align}
 In this case, the Kraus representation
 $\{F_k\}$ is the canonical Kraus representation of ${\cal E}$ (i.e., satisfying ${\rm tr}F_k^\dagger F_{k'}=0$ for any $k\neq k'$). Furthermore, let $V_k\equiv F_k/\sqrt{\lambda_k}$, then ${\cal E}$ can be written as a convex combination of orthogonal isometric channels,
\begin{align}
\mathcal{E}(\rho)=\sum_{k=1}^m \lambda_k V_k\rho V_k^\dagger,\quad \forall \ \rho\in \mathcal{S}(H_a), \label{R1}
\end{align}
where  $V_k\in L(H_a,H_b)$ are isometric operators satisfying
\begin{align}
V_k^\dagger V_{k'}=\delta_{kk'}{\bf 1}_a,  \qquad   k,k'=1,2,\cdots,m. \label{R1ISO}
\end{align}
This directly implies that $m \leq \lfloor d_b/d_a\rfloor$.

\vskip 0.2cm
 {\bf (R2)$\Rightarrow$(R0)}
 
 Suppose that (R2) holds.
  Let $P_{\perp}={\bf 1}_b-\sum_{k=1}^m  V_k V_k^\dagger$.
 Consider the map  $ \mathcal{R}: \mathcal{S}(H_b)\rightarrow \mathcal{S}(H_a)$ defined as
\begin{align}
\mathcal{R}(\sigma)\equiv \sum_{k=1}^m  V_k^\dagger \sigma  V_k+ ({\rm tr} P_{\perp}\sigma) |\psi\rangle\langle\psi|,\quad  \forall \ \sigma  \in \mathcal{S}(H_b).
\end{align}
Here $|\psi\rangle$ is a fixed quantum state on $H_a.$
 Then it is easy to verify that  $ \mathcal{R}$ is a quantum channel and that 
 \begin{align}
\mathcal{R}\circ \mathcal{E}=\mathcal{I}_a.
\end{align}
Hence, (R0) holds. 
 \vskip 0.2cm
We have now completed the proof of the equivalence between  (R0), (R1), and (R2).
In the following, we will use this equivalence to prove the remaining characterizations.
\vskip 0.2cm

 {\bf (R3)$\Leftrightarrow$(R1)}

  Suppose that (R3) holds. Then from $\mathcal{I}_a=(\mathcal{E}^\dagger\circ \mathcal{E})/\beta$, it follows that $\{E_j^\dagger E_{j'}/\sqrt{\beta}\}$ is a Kraus representation of  the identity channel $\mathcal{I}_a.$ Concequently,
\begin{align}
\frac {1}{\sqrt{\beta}}E_j^\dagger E_{j'} \propto {\bf 1}_a,
\end{align}
which implies that  (R1) holds.

For the converse, supposing that  (R1) holds, it is easy to verify that
\begin{align}  \nonumber
\mathcal{E}^\dagger\circ \mathcal{E}(\rho)=&\sum_{jj'}(E_j^\dagger E_{j'})\rho (E_j^\dagger E_{j'})^\dagger\\
=&\Big (\sum_{jj'}|\lambda_{jj'}|^2\Big)\rho.
\end{align}
Setting
$
\beta=\sum_{jj'}|\lambda_{jj'}|^2,
$ we conclude that (R3) holds. 

Next, we evaluate the explicit expressions of $\beta.$
  Applying both sides of $\mathcal{E}^\dagger\circ \mathcal{E}=\beta \,{\cal I}_a$ to ${\bf 1}_a$ and then taking the trace yields
\begin{align}
\beta=\frac {1}{d_a}{\rm tr}\big (\mathcal{E}^\dagger\circ \mathcal{E}({\bf 1}_a) \cdot{{\bf 1}_a}\big )=\frac {1}{d_a}{\rm tr}\big (\mathcal{E}({{\bf 1}_a} )\big )^2.
\end{align}
Applying both sides of ${\cal I}_a\otimes \mathcal{E}^\dagger\circ \mathcal{E}={\cal I}_a\otimes \beta \,{\cal I}_a$ to $|\Phi^+\rangle\langle\Phi^+|$,
we have 
\begin{align}
{\cal I}_a\otimes \mathcal{E}^\dagger\circ \mathcal{E}(|\Phi^+\rangle\langle\Phi^+|)=\beta |\Phi^+\rangle\langle\Phi^+|.
\end{align}
Then multiplying both sides from the right by $|\Phi^+\rangle\langle\Phi^+|$ and taking the trace yields
\begin{align}  \nonumber
\beta=&{\rm tr}\Big ({\cal I}_a\otimes \mathcal{E}^\dagger\circ \mathcal{E}(|\Phi^+\rangle\langle\Phi^+|)\cdot |\Phi^+\rangle\langle\Phi^+|\Big )\\  \nonumber
=&{\rm tr}\Big ({\cal I}_a\otimes \mathcal{E}(|\Phi^+\rangle\langle\Phi^+|)\cdot {\cal I}_a\otimes \mathcal{E}(|\Phi^+\rangle\langle\Phi^+|)\Big )\\
=&{\rm tr}\big (J_{\mathcal{E}}^2\big ).
\end{align}

 \vskip 0.2cm
  {\bf (R4)$\Leftrightarrow$(R0)}

   Suppose that (R0) holds, which implies that (R2) holds.
     Then
\begin{align}\label{channelV}
\mathcal{E}(\rho)=\sum_{k=1}^m \lambda_k V_k\rho V_k^\dagger,\quad \forall \ \rho\in \mathcal{S}(H_a),  \\  \label{dualchannelV}
\mathcal{E}^\dagger (\sigma)=\sum_{k=1}^m \lambda_k V_k^\dagger \sigma V_k,\quad \forall \ \sigma\in \mathcal{S}(H_b),
\end{align}
and
\begin{align}
\sqrt{\mathcal{E}({\bf 1}_a)}^{-1}=\sum_{k=1}^m \frac 1{\sqrt{\lambda_k}} V_k V_k^\dagger.
\end{align}
Thus, from $V_k^\dagger V_{k'}=\delta_{kk'}{\bf 1}_a$, we get that for any  $\rho\in \mathcal{S}(H_a),$ it holds that
\begin{align}
\mathcal{R}^{\rm Petz}_{{\mathcal{E}}} \circ \mathcal{E}(\rho)={\cal E}^\dagger \big (\sqrt{\mathcal{E}({\bf 1}_a)}^{-1}{\cal E}(\rho)\sqrt{\mathcal{E}({\bf 1}_a)}^{-1}\big )=\rho.
\end{align}
Hence (R4) holds.

For the converse, suppose that  (R4) holds.
Consider the map  $\mathcal{R}:\mathcal{S}(H_b)\rightarrow \mathcal{S}(H_a)$  defined as
\begin{align}
\mathcal{R}(\sigma)= \mathcal{R}^{\rm Petz}_{\mathcal{E}}(P\sigma P)+({\rm tr} P_{\perp}\sigma) |\psi\rangle\langle\psi|,\quad \forall\,  \sigma\in {\cal S}(H_b),
\end{align}
where  $P=\Pi_{\mathcal{E}({\bf 1}_a)}$, $P_\perp={\bf 1}_b-P$, and 
  $|\psi\rangle$ is a fixed quantum state on $H_a.$
It is easy to verify that ${\cal R}$ is a quantum channel. 
Then (R4) implies that ${\cal R}$ is a recovery channel of ${\cal E}$. 
 Thus, (R0) holds.
 \vskip 0.2cm

   {\bf (R5)$\Leftrightarrow$(R4)} 
   
(R5) directly implies (R4), because the rotated Petz recovery map reduces to the standard Petz recovery map at \(t=0\), i.e.,
\begin{align}
{\cal R}^{\rm Petz}_{{\cal E};0}={\cal R}^{\rm Petz}_{{\cal E}}.
\end{align}

For the converse, suppose that  (R4) holds. Then (R2) holds. Then from Eqs. (\ref{channelV}) and (\ref{dualchannelV}), we  have 
  \begin{align}
\sqrt{{\cal E}({\bf 1}_a)}^{-1}{\cal E} ({{\bf 1}_a}/{d_a} )^{{\rm i}t}={d_a}^{-{\rm i}t}\sum_{k=1}^m\lambda_k^{-\frac 1 2+{\rm i}t} V_k V_k^\dagger.
 \end{align}
So, for any state $\rho\in {\cal S}(H_a),$
  \begin{align}\nonumber
&\sqrt{{\mathcal{E}({\bf 1}_a)}}^{-1}{\cal E}({\bf 1}_a/d_a)^{{\rm i}t} {\cal E}(\rho) \,{\cal E}({\bf 1}_a/d_a)^{-{\rm i}t}\sqrt{\mathcal{E}({\bf 1}_a)}^{-1}\\
=&\sum_{k=1}^mV_k\rho V_k^\dagger,
 \end{align}
 and thus
  \begin{align}
  \mathcal{R}^{\rm Petz}_{\mathcal{E};t}\circ {\cal E}  (\rho) &= \mathcal{E}^\dagger\Big (\sum_{k=1}^mV_k\rho V_k^\dagger\Big )=\sum_{k=1}^m  \lambda_k\rho=\rho.
 \end{align}
Hence, (R5) holds.

 \vskip 0.2cm

   {\bf (R6)$\Leftrightarrow$(R4)}

 Suppose that (R6) holds. 
Let $\gamma_1={\bf 1}_a/d_a$ and $\gamma_2=\gamma$ be an arbitrary state in ${\cal S}(H_a)$, then  from the assumption it follows that 
\begin{align}
\mathcal{R}^{\rm Petz}_{\mathcal E}\circ {\cal E}(\gamma)=\mathcal{R}^{\rm Petz}_{\mathcal E,\gamma}\circ {\cal E}(\gamma).
\end{align}
On the other hand, the Petz recovery map $\mathcal{R}^{\rm Petz}_{\mathcal E,\gamma}$ is constructed so that it exactly inverts ${\cal E}$ on the support of $\gamma,$ i.e.,
\begin{align}
\mathcal{R}^{\rm Petz}_{\mathcal E,\gamma}\circ {\cal E}(\gamma)
=\gamma.
\end{align}
Combining the above two equations and using the arbitrariness of $\gamma$, we conclude that (R4)  holds.

For the converse, suppose that (R4) holds. Then the Petz recovery map $\mathcal{R}^{\rm Petz}_{\mathcal E}$  satisfies
$
\mathcal{R}^{\rm Petz}_{\mathcal E}\circ {\cal E}={\cal I}_a.
$
Hence, for any $\rho,\gamma\in {\cal S}(H_a)$, 
\begin{align}
\mathcal{R}^{\rm Petz}_{\mathcal E}\circ {\cal E}(\rho)=\rho, \quad \mathcal{R}^{\rm Petz}_{\mathcal E}\circ {\cal E}(\gamma)=\gamma.
\end{align}
When  ${\rm supp}\rho \subseteq {\rm supp}\gamma,$ applying the data-processing inequality for the quantum relative entropy to both ${\cal E}$ and $\mathcal{R}^{\rm Petz}_{\mathcal E}$, we obtain
\begin{align}
S (\rho  \|\gamma )\leq S\big (\mathcal{E}(\rho)  \|\mathcal{E}  (\gamma )\big )\leq S (\rho  \|\gamma ),
\end{align}
and therefore 
\begin{align}
S (\rho  \|\gamma )= S\big (\mathcal{E}(\rho)  \|\mathcal{E}  (\gamma )\big ).
\end{align}
By the equality condition for the data-processing inequality \cite{Petz1986,Petz1988,Petz2003},   we have 
\begin{align}
\mathcal{R}^{\rm Petz}_{\mathcal E,\gamma}\circ {\cal E}(\rho)=\rho.
\end{align}
Combining this with $\mathcal{R}^{\rm Petz}_{\mathcal E}\circ {\cal E}(\rho)=\rho,$ we obtain 
\begin{align}
\mathcal{R}^{\rm Petz}_{\mathcal E}\circ {\cal E}(\rho)=\mathcal{R}^{\rm Petz}_{\mathcal E,\gamma}\circ {\cal E}(\rho).
\end{align}
Since $\gamma$ is arbitrary, for any $\gamma_1,\gamma_2\in {\cal S}(H_a)$ and any state $\rho\in {\cal S}(H_a)$ satisfying  ${\rm supp}\rho \subseteq {\rm supp}\gamma_1 \cap  {\rm supp}\gamma_2,$ we have 
\begin{align}
\mathcal{R}^{\rm Petz}_{\mathcal E,\gamma_1}\circ {\cal E}(\rho)=\mathcal{R}^{\rm Petz}_{\mathcal E}\circ {\cal E}(\rho)=\mathcal{R}^{\rm Petz}_{\mathcal E,\gamma_2}\circ {\cal E}(\rho),
\end{align}
which proves (R6). 

This characterization shows that  every state-dependent Petz recovery map $\mathcal{R}^{\rm Petz}_{\mathcal E,\gamma}$ coincides with the state-independent Petz recovery map $\mathcal{R}^{\rm Petz}_{\mathcal E}$ on the support of the corresponding prior state $\gamma$, thereby establishing the prior-state independence of Petz recovery maps.
 \vskip  0.2cm

 {\bf (R7)$\Leftrightarrow$(R2)}

  Suppose that (R2) holds, namely,
\begin{align}
\mathcal{E}(\rho)
=
\sum_{k=1}^{m}\lambda_kV_k\rho V_k^\dagger,
\end{align}
where \(V_k^\dagger V_{k'}=\delta_{k{k'}}{\bf1}_a\). 
Let $|\Phi^+\rangle= (\sum_{i=1}^{d_a}|i\rangle\otimes |i\rangle)/\sqrt{d_a}$ be a maximally entangled state on $H_a\otimes H_a$ with $\{|i\rangle:i=1,2,\cdots,d_a\}$ an orthonormal basis of  $H_a$, and for each $k=1,2,\cdots,m,$ define
\begin{align}
|\Psi_k\rangle= ({\bf 1}_a\otimes V_k)|\Phi^+\rangle\in H_a\otimes H_b,
\end{align}
 then we directly obtain that 
\begin{align}
J_{\cal E}=\sum_{k=1}^{m}\lambda_k|\Psi_k\rangle\langle \Psi_{k}|,
\end{align}
and
\begin{align}\nonumber
{\rm tr}_b(|\Psi_k\rangle\langle\Psi_{k'}|)&=\frac {1}{d_a}\sum_{i,i'}|i\rangle\langle i'|\otimes {\rm tr}(V_k|i\rangle\langle i'|V_{k'}^\dagger)\\ \nonumber
&=\frac {1}{d_a}\sum_{i,i'}|i\rangle\langle i'|\otimes   \langle i'|V_{k'}^\dagger V_k|i\rangle\\
& = \delta_{kk'}\frac {{\bf 1}_a}{d_a}.
\end{align}
That is, (R7) holds.

For the converse, suppose that (R7) holds. Let $|\Phi^+\rangle=( \sum_{i=1}^{d_a}|i\rangle\otimes |i\rangle)/\sqrt{d_a}$ be a maximally entangled state on $H_a\otimes H_a$ with $\{|i\rangle:i=1,2,\cdots,d_a\}$ an orthonormal basis of  $H_a$. For each $k=1,\cdots, m,$ define the operator $V_k\in {\cal L}(H_a,H_b)$ by 
\begin{align} \label{Vk}
V_k|i\rangle=\sqrt{d_a} (\langle i|\otimes {\bf 1}_b)|\Psi_k\rangle\in H_b.
\end{align}
Then 
\begin{align}  \nonumber
&\sum_k\lambda_k({\bf 1}_a\otimes V_k)|\Phi^+\rangle\langle\Phi^+|({\bf 1}_a\otimes V_k^\dagger)\\  \nonumber
=&  \sum_{i,i',k}\frac {\lambda_k}{d_a}  |i\rangle\langle i'| \otimes         V_k |i\rangle\langle i'|V_k^\dagger\\  \nonumber
=&  \sum_{i,i',k} {\lambda_k}\,   |i\rangle\langle i'| \otimes     (\langle i|\otimes {\bf 1}_b)|\Psi_k\rangle \langle \Psi_k|  (| i'\rangle \otimes {\bf 1}_b)\\  \nonumber
=&  \sum_{k}  {\lambda_k}\,  \Big ( \sum_i |i\rangle\langle i| \otimes {\bf 1}_b\Big )|\Psi_k\rangle \langle \Psi_k|\Big ( \sum_{i'}| i'\rangle\langle i'| \otimes {\bf 1}_b\Big ) \\ \label{ChoisumR7}
=&J_{\cal E}.
\end{align}
Recal that Channel-state duality tells us  
\begin{align}\label{Duality}
\mathcal E(\rho) =& d_a \cdot \operatorname{tr}_a  \big ((\rho^T \otimes \mathbf{1}_b) J_{\mathcal E}\big ),
 \end{align}
 where $\rho^T = \sum_{i,i'} \langle i'|\rho|i\rangle \, |i\rangle\langle i'|$ is the transpose of $\rho$ with respect to the basis $\{|i\rangle\}$. Moreover, from Eq. (\ref{Vk}), it follows  that 
 \begin{align} \nonumber
&\operatorname{tr}_a \big ( (\rho^T \otimes \mathbf{1}_b) |\Psi_k\rangle\langle\Psi_k| \big ) \\  \nonumber
=&\operatorname{tr}_a \Big ( (\rho^T \otimes \mathbf{1}_b)\big (\sum_i |i\rangle\langle i|\otimes {\bf 1}_b \big ) |\Psi_k\rangle\langle\Psi_k| \big(\sum_{i'} |i'\rangle\langle i'|\otimes {\bf 1}_b\big)\Big ) \\  \nonumber
=& \frac{1}{d_a} \sum_{i,i'} \operatorname{tr}\big ( \rho^T |i\rangle\langle i'| \big ) \cdot V_k |i\rangle\langle i'| V_k^\dagger \\  \nonumber
=& \frac{1}{d_a} \sum_{i,i'} \langle i|\rho|i'\rangle \, V_k |i\rangle\langle i'| V_k^\dagger \\ \label{Duality2}
=& \frac{1}{d_a} V_k \rho V_k^\dagger.
\end{align}
Combining Eqs. (\ref{ChoisumR7}), (\ref{Duality}) and (\ref{Duality2}), we obtain that
\begin{align}
\mathcal E(\rho) =\sum_{k=1}^{m}\lambda_kV_k\rho V_k^\dagger. 
 \end{align}
 Then, we prove that $V_k^\dagger V_{k'}=\delta_{kk'}{\bf 1}_a.$
From the asumption that $
{\rm tr}_b(|\Psi_k\rangle\langle\Psi_{k'}|)=\delta_{kk'}{{\bf 1}_a}/{d_a}$, we get that
\begin{align} \nonumber
 \langle \Psi_k|(|i\rangle\langle i'|\otimes {\bf 1}_b)|\Psi_{k'}\rangle&={\rm tr }\big (|\Psi_{k'}\rangle\langle\Psi_k |\cdot (|i\rangle\langle i'|\otimes {\bf 1}_b)\big )\\ \nonumber
 &={\rm tr }\big ({\rm tr}_b(|\Psi_{k'}\rangle\langle\Psi_k |) \cdot |i\rangle\langle i'|\big ) \\ \nonumber
 &={\rm tr }\Big (\delta_{kk'}\frac {{\bf 1}_a}{d_a}\cdot |i\rangle\langle i'|\Big )  \\ \label{Vkk1}
 &=\frac {1}{d_a}\delta_{kk'}\delta_{ii'}.
\end{align}
From the defining equation (\ref{Vk}), we know that  $V_k$ can be written as
\begin{align} \label{Vkk2}
V_k=\sqrt{d_a}\sum_i (\langle i|\otimes {\bf 1}_b)|\Psi_k\rangle\langle i|.
\end{align}
Combining Eqs. (\ref{Vkk1}) and (\ref{Vkk2}),  we get that 
\begin{align} \nonumber
V_k^\dagger V_{k'}&=d_a\sum_i | i\rangle\langle \Psi_k|(|i\rangle\otimes {\bf 1}_b)\sum_{i'} (\langle i'|\otimes {\bf 1}_b)|\Psi_{k'}\rangle\langle i'|\\ \nonumber
&=d_a\sum_{ii'}  \langle \Psi_k|(|i\rangle\langle i'|\otimes {\bf 1}_b)|\Psi_{k'}\rangle\cdot |i\rangle\langle i'|\\ \nonumber
&=d_a\sum_{ii'} \frac {1}{d_a}\delta_{kk'}\delta_{ii'}|i\rangle\langle i'|\\
&=\delta_{kk'}{\bf 1}_a.
\end{align}
Therefore, (R2) holds.

We remark that the condition
$
{\rm tr}_b(|\Psi_k\rangle\langle\Psi_{k'}|)=\delta_{kk'}{{\bf 1}_a}/{d_a}
$
is essential. For instance, the Choi states of Pauli channels on qubit systems are Bell-diagonal states, i.e., mixtures of Bell states. Nevertheless, they generally fail to satisfy the above condition, and the corresponding channels are therefore not reversible.

Next, we prove that the spectral decomposition associated with  Eq. (\ref{Choisum}) and the  direct-sum decomposition associated with  Eq. (\ref{Choidirectsum})  are equivalent.
Suppose that the conditions associated with Eq. (\ref{Choisum})   hold. Then for each $k$, from $
{\rm tr}_b(|\Psi_k\rangle\langle\Psi_{k}|)={{\bf 1}_a}/{d_a}
$, it follows that $|\Psi_k\rangle$ is maximally entangled, namely, it can be written as
\begin{align}
|\Psi_k\rangle
=\frac 	1{\sqrt{d_a}} \sum_{i=1}^{d_a}
|i^{(k)}\rangle\otimes|i_{b_k}\rangle.
\end{align}
Here $\{|i^{(k)}\rangle:i=1,2,\cdots,d_a\}$ is an orthonormal basis of $H_a$, and 
$\{|i_{b_k}\rangle:i=1,2,\cdots,d_a\}$ is an orthonormal set in $H_b$. From 
$
{\rm tr}_b(|\Psi_k\rangle\langle\Psi_{k'}|)=0
$  for $k\neq k'$,
it follows that
\begin{align}
\langle i_{b_k}|i'_{b_{k'}}\rangle=\delta_{kk'}\delta_{ii'}.
\end{align}
Let $H_{b_k}={\rm span}\{|i_{b_k}\rangle:i=1,2,\cdots,d_a\}$ for $k=1,2,\cdots,m,$
$H_{b_0}=\big (\bigoplus_{k=1}^m H_{b_k}\big )^\perp.$
Then $H_b$ has the decomposition
\begin{align}
H_b=
\Big (\bigoplus_{k=1}^mH_{b_k}\Big )\bigoplus H_{b_0},
\qquad
H_{b_k}\cong H_a.
\end{align}
Let $\{|i\rangle:i=1,2,\cdots,d_a\}$  be the standard orthonormal basis of $H_a,$ and 
$
U^{(k)}=(u^{(k)}_{ii'})
$
denote the unitary  matrix  transforming  $\{|i\rangle:i=1,2,\cdots,d_a\}$  to   $\{|i^{(k)}\rangle:i=1,2,\cdots,d_a\}$, i.e., 
\begin{align}\nonumber
|i^{(k)}\rangle=\sum_{i'=1}^{d_a}  u^{(k)}_{ii'}|i'\rangle.
\end{align}
Noticing the fact about the maximally entangled state $  |\Phi^+\rangle\in H_a\otimes H_a$ that for any operator $X\in {\cal L}(H_a),$
\begin{align}\nonumber
X\otimes {\bf 1}_{a}  |\Phi^+\rangle=  {\bf 1}_{a} \otimes X^T  |\Phi^+\rangle,
\end{align}
let $|\tilde{i}_{b_k}\rangle=\sum_{i'} u^{(k)}_{i'i} |i'_{b_k}\rangle$ and 
$|\Phi_k^+\rangle= (\sum_{i=1}^{d_a}
|i\rangle\otimes|\tilde{i}_{b_k}\rangle)/{\sqrt{d_a}},$  then
it is easy to verify that $|\Psi_k\rangle=|\Phi_k^+\rangle\in {\cal S}(H_a\otimes H_{b_k})$ and thus
\begin{align} 
J_\mathcal{E}=\sum_{k=1}^{m}\lambda_k|\Psi_k\rangle\langle \Psi_{k}|
=
\bigoplus_{k=1}^m
\lambda_k
|\Phi_k^+\rangle\langle\Phi_k^+|.
\end{align}
That is, the conditions associated with  Eq. (\ref{Choidirectsum})  are satisfied.

For the converse,  suppose that the conditions associated with Eq. (\ref{Choidirectsum}) hold. Then, it is easy to verify that 
\begin{align}\nonumber
{\rm tr}_b(|\Phi_k^+\rangle\langle\Phi_{k'}^+|)&=\delta_{kk'}{\rm tr}_b(|\Phi_k^+\rangle\langle\Phi_{k}^+|)\\
 &=\delta_{kk'}\frac {{\bf 1}_a} {d_a}.
\end{align}
Hence, the conditions associated with  Eq. (\ref{Choisum})   are satisfied.

Now, we prove the equivalence between the direct-sum decomposition associated with  Eq. (\ref{Choitensor}) and the tensor-product decomposition associated with  Eq. (\ref{Choidirectsum}).
Suppose that the conditions associated with Eq. (\ref{Choitensor})   hold. 
Let $\tau\in {\cal S}(H_c)$ has the spectral decompositon of   $\tau=\sum_{k=1}^m\lambda_k |k_c\rangle\langle k_c|$. Without loss of generality, suppose $\lambda_k>0$, $\dim {H_c}=m$, and $\{|k_c\rangle:k=1,2,\cdots,m\}$ constitues an orthonormal basis of $H_c.$   Setting $|\Psi_k\rangle=       ({\bf 1}_a\otimes U_b)(|\Phi^+\rangle\otimes |k_c\rangle),$  we have
\begin{align} 
J_\mathcal{E}=\sum_{k=1}^{m}\lambda_k      |\Psi_k\rangle\langle \Psi_{k}|,
\end{align}
and 
\begin{align}\nonumber
{\rm tr}_b(|\Psi_k\rangle\langle\Psi_{k'}|)&= {\rm tr}_b(|\Phi^+\rangle\langle\Phi^+|\otimes |k_c\rangle\langle k'_c|)\\
 &=\delta_{kk'}\frac {{\bf 1}_a} {d_a}.
\end{align}
So,  the conditions associated with  Eq. (\ref{Choisum}) and thus the conditions associated with  Eq. (\ref{Choidirectsum})    are satisfied.

For the converse, suppose that the conditions associated with  Eq. (\ref{Choidirectsum})  holds.  under the natural identification
$\bigoplus_{k=1}^{m} H_{b_k} \cong H_a \otimes H_c$,
where $\dim H_c=m$,  let $\{|k_c\rangle\}_{k=1}^m$ be an orthonormal basis of $H_c$ and define the isomorphism
$V:H_a\otimes H_c\rightarrow \oplus_{k=1}^{m}H_{b_k}$ by
\begin{align}
V(|i\rangle \otimes |k_c\rangle)=|i_{b_k}\rangle ,
\end{align}
where $\{|i_{b_k}:i=1,2,\cdots,d_a\rangle\}$ is the basis of $H_{b_k}$ used
in the definition of $|\Phi_k^+\rangle$. Then
\begin{align}
|\Phi_k^+\rangle
=({\bf 1}_a\otimes V )
|\Phi^+\rangle\otimes |k_c\rangle.
\end{align}
Let  $\tau=\sum_{k=1}^{m}\lambda_k |k_c\rangle\langle k_c|\in\mathcal S(H_c)
$, then  
\begin{align}
J_{\mathcal E}
&=({\bf 1}_a\otimes V )(
|\Phi^+\rangle\langle\Phi^+|
\otimes \tau )({\bf 1}_a\otimes V^\dagger ).
\end{align}
 Extending  the isomorphism
$V$ to the whole output space
$H_b$ yields a unitary operator $U_b$ on $H_b$, and hence
\begin{align}
J_\mathcal{E}
=({\bf 1}_a\otimes U_b)
(|\Phi^+\rangle\langle\Phi^+|\otimes\tau)
({\bf 1}_a\otimes U_b^\dagger).
\end{align}
That is, the conditions associated with  Eq. (\ref{Choitensor})  are satisfied.

We remark that 
the tensor-product decomposition could also be derived from the statement (R19) and the saturation condition of the  Araki-Lieb inequality \cite{ArakiLieb1970,CarlenLieb2012}.

 \vskip  0.2cm
 
 {\bf (R8)$\Leftrightarrow$(R1)} 
 
 It is straightforward to see that (R8) is merely a block-matrix-form reformulation of (R1). Nevertheless, this reformulation is useful not only because it simplifies calculations in certain situations, but also because it admits a natural interpretation in terms of the Choi operator. Indeed, as shown in Eq.~(\ref{conjugateChoi}), the corresponding block operator is precisely $ d_a S_{\rm swap}\bar{J}_{\widehat{\mathcal E}}S_{\rm swap},$ where $\bar{J}_{\widehat{\mathcal E}}$  denotes  
  the complex conjugate operator  of the Choi operator ${J}_{\widehat{\mathcal E}}$ (see the derivation of Eq.~(\ref{conjugateChoi}) for details).
 
We remark that this block  operator reduces to the QEC matrix  investigated in quantum error correction setting \cite{JiangLiang2024}.
 
 \vskip  0.2cm

 {\bf (R9)$\Leftrightarrow$(R8)}
 
Given a quantum channel ${\cal E}$ with the Kraus representation $\{E_j:j=1,2,\cdots,n\}$,  
let
 \begin{align}
V=\sum_{j=1}^n E_j\otimes |j_c\rangle \in \mathcal{L}(H_a,H_b\otimes H_c).
 \end{align}
 Here  $\{|j_c\rangle:j=1,2,\cdots,n\}$ is an orthonormal basis of  $H_c \cong \mathbb{C}^n.$ Then,  $V$ is an isometry as $V^\dagger V=\sum_j E_j^\dagger E_j={\bf 1}_a.$ So,  we get a Stinespring dilation representation of $\mathcal{E}$ as
 \begin{align}
 \mathcal{E}(\rho)={\rm tr}_{c}V\rho V^\dagger,\quad \forall\, \rho\in\mathcal{S}(H_a),
 \end{align}
and  the corresponding complementary channel reads
 \begin{align}\nonumber
\widehat{\mathcal{E}}(\rho )=  {\rm tr}_{b}V\rho V^\dagger=  \sum_{j,j'=1}^n({\rm tr}E_j\rho E_{j'}^\dagger )\, |j_c\rangle\langle j'_c|, \quad \forall\,\rho\in\mathcal{S}(H_a).
 \end{align}
  Let $|\Phi^+\rangle=( \sum_{i=1}^{d_a}|i\rangle\otimes |i\rangle)/\sqrt{d_a}$ be a maximally entangled state on $H_a\otimes H_a$ with $\{|i\rangle:i=1,2,\cdots,d_a\}$ an orthonormal basis of  $H_a$, and 
  \begin{align}\label{Psi_abc}
|\Psi_{abc}\rangle =({\bf 1}_a\otimes V)|\Phi^+\rangle=\sum_{i,j}\frac {1}{\sqrt{d_a}}|i\rangle \otimes E_j|i\rangle\otimes |j_c\rangle.
\end{align}
Then, the Choi state of  $\widehat{\mathcal{E}}$ can be calculated as 
 \begin{align}  \nonumber
J_{\widehat{\cal E}}&={\rm tr}_b |\Psi_{abc}\rangle \langle\Psi_{abc}|\\  \nonumber
&=\frac {1}{{d_a}}\sum_{i,i',j,j'} {\rm tr}\big (E_j|i\rangle \langle i'| E_{j'}^\dagger\big ) \, |i\rangle \langle i'|    \otimes   |j_c\rangle\langle j'_c|\\  \nonumber
&=\frac {1}{{d_a}}\sum_{i,i',j,j'} \langle i'| E_{j'}^\dagger E_j|i\rangle  \, |i\rangle \langle i'|    \otimes   |j_c\rangle\langle j'_c|\\
&=\frac {1}{{d_a}}\sum_{j,j'} \overline{{E}_j^\dagger  { E}_{j'}} \otimes   |j_c\rangle\langle j'_c|.
\end{align}
Here $ \overline{{E}_j^\dagger  { E}_{j'}}$ denotes the  complex conjugate operator of $ {{E}_j^\dagger  { E}_{j'}}\in {\cal L}(H_a)$  with respect to the basis $\{|i\rangle\}$ of $H_a$.
Therefore,  
 \begin{align}\label{conjugateChoi}
\bar{J}_{\widehat{\cal E}}&=\frac {1}{{d_a}}\sum_{j,j'} {E}_j^\dagger   { E}_{j'} \otimes   |j_c\rangle\langle j'_c|,
\end{align}
where $\bar{J}_{\widehat{\cal E}}$ denotes the complex conjugate operator of ${J}_{\widehat{\cal E}}$ with respect to the basis $\{|i\rangle \otimes | j_c\rangle\}$ of $H_a\otimes H_b.$
(R9)$\Leftrightarrow$(R8) immediately follows from  the fact that 
 $J_{\widehat{\cal E}}$ is a product state if and only if 
 \begin{align}
  \sum_{j,j'}|j_c\rangle\langle j'_c|\otimes E_j^\dagger E_{j'}=d_a S_{\rm swap}\bar{J}_{\widehat{\mathcal E}}S_{\rm swap}
\end{align} 
  is of product form. Here $S_{\rm swap}=\sum_{ij}|j_c\rangle\langle i|\otimes |i \rangle\langle j_c|$ denotes the swap operator on $H_a\otimes H_c$.

\vskip 0.2cm
 {\bf (R10)$\Leftrightarrow$(R9)}
 
Suppose that (R10) holds. By defintion, we have
\begin{align}  \nonumber
J_{\widehat{\cal E}}&={\cal I}_a\otimes \widehat{\cal E}(|\Phi^+\rangle\langle\Phi^+|)\\  \nonumber
&=\frac{1}{d_a}\sum_{i,i'}|i\rangle\langle i'|\otimes \widehat{\cal E}(|i\rangle\langle i'|)\\  \nonumber
&=\frac {1}{d_a}\sum_{i,i'}|i\rangle\langle i'|\otimes  {\rm tr}(|i\rangle\langle i'|) \tau\\
&=\frac {{\bf 1}_a}{d_a}\otimes   \tau.
\end{align}
(R9) directly follows from $\widehat{\cal E}({\bf 1}_a/d_a)=\tau$.

For the converse, suppose that (R9) holds.   
Recal that Channel-state duality tells us  
\begin{align}
\widehat{\cal E}(X) =& d_a \cdot \operatorname{tr}_a  \big ((X^T \otimes \mathbf{1}_b) J_{\widehat{\cal E}}\big ),
 \end{align}
 where $X^T = \sum_{i,i'} \langle i'|X|i\rangle \, |i\rangle\langle i'|$ is the transpose of $X$ with respect to the basis $\{|i\rangle\}$.  Substituting Eq. (\ref{R9}) in  (R9) into the above equation, we get that 
 \begin{align} \nonumber
\widehat{\cal E}(X)&= d_a \cdot \operatorname{tr}_a  \Big ((X^T \otimes \mathbf{1}_b)  \Big ( \frac {{\bf 1}_a}{d_a}\otimes \widehat{\cal E}\Big (\frac {{\bf 1}_a}{d_a}
   \Big)  \Big)  \Big ),   \\  \nonumber
&={\rm tr}_a \Big (X^T\otimes \widehat{\cal E}\Big (\frac {{\bf 1}_a}{d_a}\Big)\Big),\\
&=({\rm tr}X)\, \widehat{\cal E}\Big (\frac {{\bf 1}_a}{d_a}\Big),
\end{align}
which implies that (R10) holds.

\vskip 0.2cm
 {\bf (R11)$\Leftrightarrow$(R9)}

Let   $\{|i\rangle:i=1,2,\cdots,d_a\}$ be   an orthonormal basis of  $H_a$ and $H_{a'}= H_a$ with the same basis, then the Jordan product of  $\mathcal{I}_a$ and $\widehat{\mathcal{E}}$ is
\begin{align}
C_{\mathcal{I}_a\circ\widehat{\mathcal{E}}}= \sum_{i,i',j,j'}\{|i\rangle\langle i'|, |j\rangle\langle j'|\}\otimes |i\rangle\langle i'|\otimes \widehat{\mathcal{E}}(|j\rangle\langle j'|).
\end{align}
with $C_{\mathcal{I}_a\circ\widehat{\mathcal{E}}} \in {\cal L}(H_a\otimes H_{a'}\otimes H_c).$
Suppose that  (R11) holds, namely,  $C_{\mathcal{I}_a\circ\widehat{\mathcal{E}}}\geq 0$. 
Since
\begin{align}
{\rm tr}_{c}C_{\mathcal{I}_a\circ\widehat{\mathcal{E}}}&=  d_a\, |\Phi^+\rangle\langle \Phi^+|
\end{align}
is rank one,  
using the fact that a positive semidefinite bipartite operator with a rank-one marginal is necessarily a product operator, we have
\begin{align}
C_{\mathcal{I}_a\circ\widehat{\mathcal{E}}}= d_a|\Phi^+\rangle\langle \Phi^+|\otimes \tau
\end{align}
 with $\tau$ being a density operator on $H_c$ (i.e., $\tau\geq 0$ and  ${\rm tr}\tau=1$). 
   In this case,
 \begin{align}
C_{\widehat{\mathcal{E}}}={\rm tr}_{a'}C_{\mathcal{I}_a\circ\widehat{\mathcal{E}}}&= {\bf 1}_a  \otimes \tau,
\end{align}
 which implies (R9) holds.

For the converse, suppose that (R9) holds. Then for $ i = i'$, we have  
$\widehat{\cal E}(|i\rangle\langle i'|) = \widehat{\cal E}\big ( {{\bf 1}_a}/{d_a}\big)$; for $i\neq i',$ $\widehat{\cal E}(|i\rangle\langle i'|) =0.$ 
Direct calculation shows that 
\begin{align}\nonumber
C_{\mathcal{I}_a\circ\widehat{\mathcal{E}}}&= \sum_{i,i'}|i\rangle\langle i'|\otimes |i\rangle\langle i'|\otimes \widehat{\cal E}\Big (\frac {{\bf 1}_a}{d_a}\Big)\\
&= d_a|\Phi^+\rangle\langle \Phi^+|\otimes \widehat{\cal E}\Big (\frac {{\bf 1}_a}{d_a}\Big)\geq 0.
\end{align}
That is, (R11) holds.
 \vskip 0.2cm
{\bf   (R12)$\Leftrightarrow$(R0)}

  Suppose that (R0) holds, which implies that (R2) holds. Let $H_{c}\cong \mathbb{C}^m$ with an orthonormal basis $\{k_{c}\rangle:k=1,2,\cdots,m\}$, and $\tau=\sum_{k=1}^m\lambda_k|k_{c}\rangle\langle k_{c}|\in {\cal S}(H_{c})$. Define $W=\sum_kV_k\otimes \langle k_{c}|\in \mathcal{L}(H_a\otimes H_{c},H_b)$.
  Since $V_k^\dagger V_{k'}=\delta_{kk'}{\bf 1}_a$, we have 
\begin{align}
W^\dagger  W={\bf 1}_{a}\otimes {\bf  1}_{c}.
\end{align}
Hence, $W$ is an isometry. Moreover, for any state $\rho\in {\cal S}(H_a),$
\begin{align}\nonumber
W(\rho\otimes\tau)W^\dagger
&=
\sum_{k=1}^m
\lambda_k
W(\rho\otimes|k_{c}\rangle\langle k_{c}|)W^\dagger \\ \label{Wtau}
&=
\sum_{k=1}^m
\lambda_k
V_k\rho V_k^\dagger
=
\mathcal E(\rho).
	\end{align}
That is, (R12) holds.

 For the converse, suppose that (R12) holds. That is,  $\mathcal{E}(\rho)=W(\rho\otimes\tau)W^\dagger$.
Consider the map  $\mathcal{R}:\mathcal{S}(H_b)\rightarrow \mathcal{S}(H_a)$, which is defined as
\begin{align}
\mathcal{R}(\sigma)={\rm tr}_{c}\big( W^\dagger \sigma W\big)+{\rm tr}(P_\perp \sigma )|\psi\rangle\langle \psi|,\quad \forall \  \sigma \in \mathcal{S}(H_b),
\end{align}
where $P=WW^\dagger$, $P_\perp={\bf 1}_b-P$, and $|\psi\rangle$ is a fixed state on $H_a$.
Then it is easy to verity that ${\cal R}$ is a quantum channel and   ${\cal R}\circ \mathcal{E}={\cal I}_a,$ and thus (R0) holds.

Note that the equivalence between Eqs.~(\ref{Isodecomp}) and (\ref{Unidecomp})
follows directly from the fact that every isometry can be extended to a
unitary operator on the output space.

\vskip 0.2cm
 We postpone the proof of  (R13)$\Leftrightarrow$(R0) until  we have proved  (R14)$\Leftrightarrow$(R0), as  we will use the equivalence between (R14) and (R0) to prove that between (R13) and (R0).

\vskip 0.2cm
{\bf (R14)$\Leftrightarrow$(R0)}

   Suppose that (R0) holds, which implies that (R2) holds. Hence,
 for any  $\rho,\sigma\in \mathcal{S}(H_a)$,
\begin{align}
\mathcal{E}(\rho\sigma)\mathcal{E}({\bf 1}_a)&=\sum_{i=1}^m\lambda_i V_i\rho\sigma V_i^\dagger \sum_{j=1}^m\lambda_j V_j{\bf 1}_a V_j^\dagger  \nonumber \\
&=\sum_{i=1}^m\lambda_i^2 V_i\rho\sigma V_i^\dagger  \nonumber \\
&=\mathcal{E}(\rho)\mathcal{E}(\sigma).
\end{align}

For the converse, suppose that (R14) holds.
Due to the cyclic property of the trace, we have
\begin{align}
&{\rm tr}\big (\mathcal{E}(\rho\sigma)\mathcal{E}({\bf 1}_a)\big )={\rm tr}\big ({\cal E}^\dagger\circ\mathcal{E}
({\bf 1}_a)\rho\sigma\big )={\rm tr}\big (\sigma ({\cal E}^\dagger\circ\mathcal{E})({\bf 1}_a)\rho \big ),\\
&{\rm tr}\big (\mathcal{E}(\rho)\mathcal{E}(\sigma)\big )={\rm tr}\big ({\cal E}^\dagger\circ\mathcal{E}(\rho)\sigma\big )={\rm tr}\big ({\cal E}^\dagger\circ\mathcal{E}(\sigma)\rho\big ),
\end{align}
for any $\rho, \sigma\in \mathcal{S}(H_a).$
Combining the above equations and the assumption, we obtain that for a fixed state $\sigma\in \mathcal{S}(H_a)$,
\begin{align}
{\rm tr}\big (\sigma{\cal E}^\dagger\circ\mathcal{E}({\bf 1}_a)\rho\big )={\rm tr}\big ({\cal E}^\dagger\circ\mathcal{E}(\sigma)\rho\big ), \quad \forall \  \rho\in \mathcal{S}(H_a),
\end{align}
which implies that
\begin{align}\label{v1}
\sigma{\cal E}^\dagger\circ\mathcal{E}({\bf 1}_a)={\cal E}^\dagger\circ\mathcal{E}(\sigma), \quad  \forall \ \sigma\in \mathcal{S}(H_a).
\end{align}
Similarly, we get that
for a fixed state $\rho\in \mathcal{S}(H_a)$,
\begin{align}
{\rm tr}\big ({\cal E}^\dagger\circ\mathcal{E}({\bf 1}_a)\rho\sigma\big )={\rm tr}\big ({\cal E}^\dagger\circ\mathcal{E}(\rho)\sigma\big ), \quad \forall \  \sigma\in \mathcal{S}(H_a),
\end{align}
which implies that for any state $\rho\in \mathcal{S}(H_a)$,
\begin{align}\label{v2}
{\cal E}^\dagger\circ\mathcal{E}({\bf 1}_a)\rho={\cal E}^\dagger\circ\mathcal{E}(\rho).
\end{align}
Combining Eqs. (\ref{v1}) and (\ref{v2}), we obtain that for any state $\rho\in \mathcal{S}(H_a)$,
\begin{align}
{\cal E}^\dagger\circ\mathcal{E}({\bf 1}_a)\rho=\rho\,{\cal E}^\dagger\circ\mathcal{E}({\bf 1}_a)={\cal E}^\dagger\circ\mathcal{E}(\rho).
\end{align}
Due to the arbitrariness of $\rho$, there exists a constant $\beta$ such that
\begin{align}
{\cal E}^\dagger\circ\mathcal{E}({\bf 1}_a)=\beta {\bf 1}_a,
\end{align}
and thus
\begin{align}
 {\cal E}^\dagger\circ\mathcal{E}(\rho)={\beta }\rho.
\end{align}
So  (R3) holds, which implies that (R0) holds.
This completes the proof of (R14)$\Leftrightarrow$(R0).

\vskip 0.2cm
  {\bf (R13)$\Leftrightarrow$(R0)}

Suppose that (R13) holds. Then for any state $\rho, \sigma\in \mathcal{S}(H_a),$
\begin{align}\nonumber
\mathcal{E}(\rho)\mathcal{E}(\sigma)&=\pi(\rho)\mathcal{E}({\bf 1}_a)\pi(\sigma)\mathcal{E}({\bf 1}_a)\\ \nonumber
&=\pi(\rho)\pi(\sigma)\mathcal{E}({\bf 1}_a)\mathcal{E}({\bf 1}_a)\\ \nonumber
&=\pi(\rho\sigma)\mathcal{E}({\bf 1}_a)\mathcal{E}({\bf 1}_a)\\
&=\mathcal{E}(\rho\sigma)\mathcal{E}({\bf 1}_a).
\end{align}
The first and last equalities use ${\cal E}(X)=\pi(X){\cal E}({\bf 1}_a)$, the second uses the commutativity $\pi(X){\cal E}({\bf 1}_a)={\cal E}({\bf 1}_a)\pi(X)$,  and the third uses the multiplicativity of $\pi.$ 
Similarly,  
\begin{align}
\mathcal{E}(\rho)\mathcal{E}(\sigma)&=\mathcal{E}({\bf 1}_a)\mathcal{E}(\rho\sigma).
\end{align}
Therefore,  (R14) holds, and thus (R0) holds. 

For the converse, suppose that (R0) holds, which implies that (R2) holds. That is, $\mathcal{E}$ can be expressed as a convex combination of orthogonal isometric channels
\begin{align}
\mathcal{E}(\rho)=\sum_{k=1}^m \lambda_k V_k\rho V_k^\dagger,\quad \forall \ \rho\in \mathcal{S}(H_a).
\end{align}
Then
$
\mathcal{E}({\bf 1}_a)=\sum_{k=1}^m \lambda_k V_kV_k^\dagger.
$
Define $\pi:{\cal L}(H_a)\rightarrow {\cal L}(\Pi_{{\cal E}({\bf 1}_b)}H_b)$ by
\begin{align}
\pi(X)=\sum_{k=1}^m   V_kXV_k^\dagger,\quad\quad \forall \ X\in \mathcal{L}(H_a).
\end{align}
Obviously, $\pi$ is linear and  preserves multiplication, i.e., for any $X,Y\in {\cal L}(H_a),$
\begin{align}
\pi(X)\pi(Y)=\sum_{k=1}^m   V_kXYV_k^\dagger=\pi(XY).
\end{align}
Furthermore,
\begin{align}
\pi(X^\dagger)=\sum_{k=1}^m   V_kX^\dagger V_k^\dagger=\pi(X)^\dagger.
\end{align}
Thus, $\pi$ is a representation and it is easy to vefiry that
\begin{align}
\mathcal{E}(X)=\pi (X) \mathcal{E}({\bf 1}_a)= \mathcal{E}({\bf 1}_a)\pi (X),\quad \forall \, X\in \mathcal{L}(H_a).
\end{align}
 This implies that (R13) holds. This completes the proof of
(R13)$\Leftrightarrow$(R0).

\vskip 0.2cm
{\bf (R15)$\Leftrightarrow$(R0)}

 Suppose that (R0) holds, which implies that (R14) holds. Then it is easy to get that  (R15) holds.
 For the converse,  suppose that (R15) holds. Let $|\psi\rangle$ and $|\varphi\rangle$ be two orthogonal pure states on $H_a,$ then
 \begin{align}
 \mathcal{E}(|\psi\rangle\langle\psi|)\mathcal{E}(|\varphi\rangle\langle\varphi|)=0.
 \end{align}
Taking the trace on both sides of the above equation, we obtain that for any Kraus representation $\{E_j\}$ of $\mathcal{E}$, it holds that
 \begin{align}
\sum_{jj'}  | \langle\psi|E_j^\dag E_{j'}|\varphi \rangle|^2=0,
 \end{align}
 which implies that
 \begin{align}
  \langle\psi|E_j^\dag E_{j'}|\varphi \rangle=0, \qquad  \forall \ j, j'
 \end{align}
 and thus
 \begin{align}
|\psi\rangle  \langle\psi|E_j^\dag E_{j'}|\varphi\rangle  \langle\varphi |=0.
 \end{align}
Let $P_{\psi}=|\psi\rangle\langle\psi|$, then from the arbitrariness of $|\varphi \rangle$, it follows that
 \begin{align}
P_{\psi}E_j^\dag E_{j'}({\bf 1}_a-P_{\psi})=0.
 \end{align}
Thus,
 \begin{align}
P_{\psi}E_j^\dag E_{j'}=P_{\psi}E_j^\dag E_{j'}P_{\psi}.
 \end{align}
 Similarly, we have
 \begin{align}
P_{\psi}E_j^\dag E_{j'}P_{\psi}=E_j^\dag E_{j'}P_{\psi}.
 \end{align}
 Hence,
 \begin{align}
P_{\psi}E_j^\dag E_{j'}=E_j^\dag E_{j'}P_{\psi}.
 \end{align}
 From the arbitrariness of $|\psi\rangle$, it follows that
 \begin{align}
E_j^\dagger E_{j'} \propto {\bf 1}_a, \qquad \forall \ j,j'
\end{align}
which implies that  (R1) holds, and thus (R0) holds.

  \vskip 0.2cm
{\bf (R16)$\Leftrightarrow$(R0)}

 Suppose that (R0) holds, which implies that (R14) holds. Thus, (R16) holds.
 For the converse, suppose that (R16) holds, which implies that (R15) holds. From   (R15)$\Leftrightarrow$(R0), it follows that (R0) holds.

\vskip 0.2cm

{\bf (R17)$\Leftrightarrow$(R0)}

  Suppose that (R0) holds, which implies that (R12) holds. Using Eq.  (\ref{Unidecomp}) and the unitary invariance of Schatten $p$-norms, we know that 
  for any $p\geq 1,$ and any  $ X \in {\cal L}(H_a),$
   \begin{align}
\|{\cal E}(X)\|_p=\|(X\otimes \tau)\oplus {\bf 0}_{b_0}\|_p=\|\tau\|_p\|X\|_p.
\end{align}
Hence, for every $p\geq 1$, there exists a number $\gamma_p=\|\tau\|_p\in(0,1]$ such that
\begin{align}
\|\mathcal{E}(X)\|_p=\gamma_p \|X\|_p,\quad \forall\ X \in {\cal L}(H_a).
\end{align}

For the converse, suppose that (R17) holds.
There exist some $p\geq1$ and a corresponding constant 
$\gamma_p\in(0,1]$ such that
\begin{align}
\|\mathcal{E}(X)\|_p
=
\gamma_p\|X\|_p,
\qquad \forall\,X\in\mathcal{L}(H_a).
\end{align}
 Let $|\varphi_1\rangle$ and $|\varphi_2\rangle$ be two arbitrary orthogonal pure states on $H_a,$ then  
  \begin{align}
  \||\varphi_1\rangle\langle\varphi_1|-|\varphi_2\rangle\langle\varphi_2|\|^p_p=2,
\end{align}
and 
\begin{align}
\|\mathcal{E}(|\varphi_i\rangle\langle\varphi_i|)\|^p_p=\gamma_p^p\||\varphi_i\rangle\langle\varphi_i|\|^p_p=\gamma_p^p,\quad {\rm for}\quad i=1,2.
\end{align}
Consequently,
\begin{align}\nonumber
&\|\mathcal{E}(|\varphi_1\rangle\langle\varphi_1|)-\mathcal{E}(|\varphi_2\rangle\langle\varphi_2|)\|^p_p\\ \nonumber
=&\|\mathcal{E}(|\varphi_1\rangle\langle\varphi_1|-|\varphi_2\rangle\langle\varphi_2|)\|^p_p
\\ \nonumber
=&\gamma^p\||\varphi_1\rangle\langle\varphi_1|-|\varphi_2\rangle\langle\varphi_2|\|^p_p\\ \nonumber
=&2\gamma_p^p\\
=&\|\mathcal{E}(|\varphi_1\rangle\langle\varphi_1|)\|^p_p+\|\mathcal{E}(|\varphi_2\rangle\langle\varphi_2|)\|^p_p.
\end{align}
Since for any positive operators $X,Y\in {\cal L}(H_a)$ and $p\geq1$, it holds that \cite{LiYuan2026}
\begin{align}
\|X-Y\|^p_p=\|X\|^p_p+\|Y\|^p_p \quad {\rm if}\  {\rm and} \ {\rm only}\  {\rm if}\quad  XY=0,
\end{align}
we conclude that
\begin{align}
\mathcal{E}(|\varphi_1\rangle\langle\varphi_1|)\mathcal{E}(|\varphi_2\rangle\langle\varphi_2|)=0.
\end{align}
 Following (R15)$\Leftrightarrow$(R0), we get that  (R17) holds.

\vskip 0.2cm
 {\bf  (R18)$\Leftrightarrow$(R0)}

 Let $D(\rho\|\sigma)$ be an arbitrary  state distingushability measure that is monotone under quantum channels (data‑processing inequality) and for which saturation of the inequality implies the existence of a recovery map.
 (R0)$\Rightarrow$(R18) can be derived  directly from  the monotonicity of $D$ and the existence of the recovery channel.

For the converse, suppose that (R18) holds. Then,
\begin{align}
D\Big (\mathcal{E}(\rho)\Big \|\mathcal{E}\Big (\frac{{\bf 1}_a}{d_a}\Big )\Big )=D\Big (\rho\Big \|\frac{{\bf 1}_a}{d_a}\Big ),\quad \quad \forall \ \rho \in \mathcal{S}(H_a).
\end{align}
From the celebrated Petz's results \cite{Petz1986,Petz1988,Petz2003}, we know that the Petz recovery channel $ \mathcal{R}^{\rm Petz}_{\cal E}:\mathcal{S}(H_b)\rightarrow \mathcal{S}(H_a)$   on the support of $\mathcal{E}  ( {{\bf 1}_a}  )$ is  determined by
\begin{align}
 \mathcal{R}^{\rm Petz}_{\cal E}(\sigma)=\mathcal{E}^\dagger  \big (  {\sqrt{ \mathcal{E}({\bf 1}_a)}}^{-1}\sigma{\sqrt{ \mathcal{E}({\bf 1}_a )}}^{-1} \big ),\quad  \forall \  \sigma \in \mathcal{S}(PH_b).
\end{align}
Here  $P=\Pi_{{\cal E}({\bf 1}_a)}$.
Furthermore,    it satisfies that
\begin{align}
 \mathcal{R}^{\rm Petz}_{\cal E}\circ \mathcal{E}(\rho)= \mathcal{R}^{\rm Petz}_{\cal E}(\mathcal{E}(\rho))=\rho,\quad  \forall \  \rho \in \mathcal{S}(H_a).
\end{align}
Therefore, let  $\mathcal {R}$  be defined as
\begin{align}
\mathcal{R}(\sigma)& = \mathcal{R}^{\rm Petz}_{\cal E}(P\sigma P)+({\rm tr} P_{\perp}\sigma) |\psi\rangle\langle\psi|,\quad \forall\,  \sigma\in {\cal S}(H_b),
\end{align}
with $P_\perp={\bf 1}_b-P$ and $|\psi\rangle$ a fixed state on $H_a,$ 
then ${\cal R}$ is a recovery channel of ${\cal E}$. Thus, (R0) holds.

Here, we provide references  establishing that the following measures satisfy the above two required properties.
For the quantum relative entropy $S(\rho\|\sigma)$, see  \cite{Petz1986,Petz1988}. For the standard R\'{e}nyi relative entropy $D_\alpha(\rho\|\sigma)$, the case $\alpha=1/2$ is covered in  Refs. \cite{Petz1986,Petz1988}, and  the remaining $\alpha$ in $(0,1)\cup(1,2)$  are treated in Refs. \cite{JencovaPetz2006a,JencovaPetz2006b}.
For the sandwiched R\'{e}nyi relative entropy  $\widetilde{D}_\alpha(\rho\|\sigma)$, the range $\alpha\in (1/2,1)$ is proven  in Refs.  \cite{Jencova2017,Jencova2018}, and  $\alpha\in (1,+\infty)$ in  \cite{Jencova2021}. For  the Chernoff distance $C(\rho\|\sigma)$, or  the Hoeffding distance $H_r(\rho\|\sigma)$ for all $r\in \mathbb{R}$, the proofs are given in \cite{Jencova2012}.
\vskip 0.2cm

  {\bf (R19)$\Leftrightarrow$(R0)}

    Similar to the above proof,  (R0)$\Rightarrow$(R19) is obvious.
    For the converse, suppose that (R19) holds, then
\begin{align}
S\Big (J_\mathcal{E} \Big \|\frac{{\bf 1}_a}{d_a}\otimes \mathcal{E}\Big (\frac{{\bf 1}_a}{d_a}\Big )\Big )=S\Big (|\Phi^+\rangle\langle\Phi^+|\Big \|\frac{{\bf 1}_a}{d_a}\otimes \frac{{\bf 1}_a}{d_a}\Big ).
\end{align}
From  Petz's results \cite{Petz1986,Petz1988,Petz2003}, we know that there exists a recovery channel $\mathcal{R}_{ab}:\mathcal{S}(H_a\otimes H_b)\rightarrow \mathcal{S}(H_a\otimes H_a)$, whose action on the support of $ {{\bf 1}_a}/{d_a}\otimes \mathcal{E}\big ( {{\bf 1}_a}/{d_a}\big )$ is uniquely determined in the sense that $ \forall \  \sigma \in \mathcal{S}(H_a\otimes \Pi_{{\cal E}({\bf 1}_a)}H_b),$
\begin{align}
\mathcal{R}_{ab}(\sigma)=\mathcal{I}_a\otimes \mathcal{E}^\dagger  \big (  {\bf 1}_a \otimes {\sqrt{ \mathcal{E}({\bf 1}_a)}}^{-1}\sigma{\bf 1}_a \otimes{\sqrt{ \mathcal{E}({\bf 1}_a )}}^{-1} \big ).
\end{align}
Note that for any Kraus representation $\{E_j\}$ of $\mathcal{E}$,
$\big \{{\bf 1}_a\otimes \big (E_j^\dagger {\sqrt{ \mathcal{E}({\bf 1}_a)}}^{-1}\big )\big \} $ forms a  Kraus representation of   $\mathcal{R}_{ab}$, when  $\mathcal{R}_{ab}$ is restricted to act on $H_a\otimes \Pi_{{\cal E}({\bf 1}_a)}H_b$.  Actually, it is easy to verify that
\begin{align}
\mathcal{R}_{ab}= \mathcal{I}_a\otimes  \mathcal{R}^{\rm Petz}_{\cal E},
 \end{align}
and
\begin{align}
 (\mathcal{I}_a\otimes \mathcal{R}^{\rm Petz}_{\cal E}\circ \mathcal{E})(|\Phi^+\rangle\langle\Phi^+| )= |\Phi^+\rangle\langle\Phi^+|,
\end{align}
which implies that
$
 \mathcal{R}^{\rm Petz}_{\cal E}\circ \mathcal{E}=\mathcal{I}_{a}.
$
Therefore, (R0) holds.

    \vskip 0.2cm

  {\bf (R20)$\Leftrightarrow$(R19)}
  
 Following the derivation and the notations in {  (R9)$\Leftrightarrow$(R8)}, we know that for  an arbitrary  quantum channel ${\cal E}$,   
 let $\sigma^{abc}=|\Psi_{abc}\rangle \langle \Psi_{abc}|$ be the pure tripartite state defined by Eq.  (\ref{Psi_abc}). Then 
\begin{align}
J_{\cal E}&=\sigma^{ab}={\rm tr}_c\sigma^{abc},\\
J_{\widehat{\cal E}}&=\sigma^{ac}={\rm tr}_b\sigma^{abc}.
\end{align}
It is easy to verify that 
 \begin{align}
S(J_{\cal E})=S(\sigma^{ab})=S(\sigma^{c}),\\
S(J_{\widehat{\cal E}})=S(\sigma^{ac})=S(\sigma^{b}),
\end{align}
which derictly implies that \cite{luo2011decoherence}
 \begin{align}\label{Informationconservation} 
I(|\Phi^+\rangle\langle\Phi^+| )=I(J_{\widehat{\cal E}})+I(J_{\cal E}).
\end{align}
So, the equivalence of (R20)$\Leftrightarrow$(R19) is straightforward from 
\begin{align}
I(J_{\widehat{\cal E}})=I(|\Phi^+\rangle\langle\Phi^+| )-I(J_{\cal E}).
\end{align}

We remark that Eq. (\ref{Informationconservation}) holds for an arbitrary quantum channel and can be interpreted as an information conservation law. For any quantum channel, the total correlations, quantified by the quantum mutual information, are decomposed into two parts: the correlations preserved in the Choi state $J_{\cal E}$ of the channel and those appearing in the Choi state $J_{\widehat{\cal E}}$ of its complementary channel.
For a reversible quantum channel, all the total correlations are preserved in $J_{\cal E}$, while none are transferred to $J_{\widehat{\cal E}}$. Consequently, perfect reversibility of the channel is exactly accompanied by maximal irreversibility of the complementary channel, which reduces to a replacement channel.

 \vskip  0.2cm

  {\bf (R21)$\Leftrightarrow$(R19)}

The equivalence is clear from
\begin{align}
S\Big (\frac {{\bf 1}_a}{d_a}\Big )-I_c\Big (\frac {{\bf 1}_a}{d_a}, {\cal E}\Big )=I(|\Phi^+\rangle\langle\Phi^+| )-I(J_{\cal E}).
\end{align}


\vskip  0.2cm

 



\begin{thebibliography}{}

\bibitem{Nielsen2012}
M. A. Nielsen and I. L. Chuang, \textit{Quantum Computation and Quantum Information} (Cambridge University Press, Cambridge, 2010).

\bibitem{Choi1975}
M.-D. Choi, Completely positive linear maps on complex matrices, Linear Algebra Appl. \textbf{10}, 285 (1975).

\bibitem{Kraus1983}
K. Kraus, \textit{States, Effects, and Operations: Fundamental Notions of Quantum Theory} (Springer, Berlin, 1983).

\bibitem{Lindblad1975}
G. Lindblad, Completely positive maps and entropy inequalities, Commun. Math. Phys. \textbf{40}, 147 (1975).

\bibitem{Uhlmann1977}
A. Uhlmann, Relative entropy and the Wigner-Yanase-Dyson-Lieb concavity in an interpolation theory, Commun. Math. Phys. \textbf{54}, 21 (1977).

\bibitem{Petz1986}
D. Petz, Sufficient subalgebras and the relative entropy of states of a von Neumann algebra, Commun. Math. Phys. \textbf{105}, 123 (1986).

\bibitem{Petz1988}
D. Petz, Sufficiency of channels over von Neumann algebras, Q. J. Math. \textbf{39}, 97 (1988).

\bibitem{Nielsen1997}
M. A. Nielsen and C. M. Caves, Reversible quantum operations and their application to teleportation, Phys. Rev. A \textbf{55}, 2547 (1997).

\bibitem{Schumacher1996}
B. Schumacher and M. A. Nielsen, Quantum data processing and error correction, Phys. Rev. A \textbf{54}, 2629 (1996).

\bibitem{Nielsen1998}
M. A. Nielsen, C. M. Caves, B. Schumacher, and H. Barnum, Information-theoretic approach to quantum error correction and reversible measurement, Proc. R. Soc. A \textbf{454}, 277 (1998).

\bibitem{Caves1999}
C. M. Caves, Quantum error correction and reversible operations, J. Supercond. \textbf{12}, 707 (1999).

\bibitem{Knill1997}
E. Knill and R. Laflamme, Theory of quantum error-correcting codes, Phys. Rev. A \textbf{55}, 900 (1997).

\bibitem{Knill2000}
E. Knill, R. Laflamme, and L. Viola, Theory of quantum error correction for general noise, Phys. Rev. Lett. \textbf{84}, 2525 (2000).

\bibitem{Ogawa2005}
T. Ogawa, A. Sasaki, M. Iwamoto, and H. Yamamoto, Quantum secret sharing schemes and reversibility of quantum operations, Phys. Rev. A \textbf{72}, 032318 (2005).

\bibitem{thermo0}
H. Wilming, R. Gallego, and J. Eisert, Axiomatic characterization of the quantum relative entropy and free energy, Entropy \textbf{19}, 241 (2017).

\bibitem{thermo1}
\'A. M. Alhambra, S. Wehner, M. M. Wilde, and M. P. Woods, Work and reversibility in quantum thermodynamics, Phys. Rev. A \textbf{97}, 062114 (2018).

\bibitem{Characterizingirreversibility2018}
T. B. Batalhão, S. Gherardini, J. P. Santos, G. T. Landi, and M. Paternostro, Characterizing irreversibility in open quantum systems, in \textit{Thermodynamics in the Quantum Regime: Fundamental Aspects and New Directions} (Springer, 2019), pp. 395-410.

\bibitem{Hiai2011}
F. Hiai, M. Mosonyi, D. Petz, and C. Bény, Quantum $f$-divergences and error correction, Rev. Math. Phys. \textbf{23}, 691 (2011).

\bibitem{Hiai2017}
F. Hiai and M. Mosonyi, Different quantum $f$-divergences and the reversibility of quantum operations, Rev. Math. Phys. \textbf{29}, 1750023 (2017).

\bibitem{Hiai2021}
F. Hiai, \textit{Quantum $f$-Divergences in von Neumann Algebras} (Springer, Singapore, 2021).

\bibitem{Ruskai2002}
M. B. Ruskai, Inequalities for quantum entropy: A review with conditions for equality, J. Math. Phys. \textbf{43}, 4358 (2002).

\bibitem{Petz2003}
D. Petz, Monotonicity of quantum relative entropy revisited, Rev. Math. Phys. \textbf{15}, 79 (2003).

\bibitem{Petz2004}
M. Mosonyi and D. Petz, Structure of sufficient quantum coarse-grainings, Lett. Math. Phys. \textbf{68}, 19 (2004).

\bibitem{Hayden2004}
P. Hayden, R. Jozsa, D. Petz, and A. Winter, Structure of states which satisfy strong subadditivity of quantum entropy with equality, Commun. Math. Phys. \textbf{246}, 359 (2004).

\bibitem{Jencova2012}
A. Jenčová, Reversibility conditions for quantum operations, Rev. Math. Phys. \textbf{24}, 1250016 (2012).

\bibitem{Shirokov2013}
M. E. Shirokov, Reversibility conditions for quantum channels and their applications, Sb. Math. \textbf{204}, 1215 (2013).

\bibitem{Petz1986quasi}
D. Petz, Quasi-entropies for finite quantum systems, Rep. Math. Phys. \textbf{23}, 57 (1986).

\bibitem{JencovaPetz2006a}
A. Jenčová and D. Petz, Sufficiency in quantum statistical inference, Commun. Math. Phys. \textbf{263}, 259 (2006).

\bibitem{JencovaPetz2006b}
A. Jenčová and D. Petz, Sufficiency in quantum statistical inference: A survey with examples, Infin. Dimens. Anal. Quantum Probab. Relat. Top. \textbf{9}, 331 (2006).

\bibitem{Jencova2017}
A. Jenčová, Preservation of a quantum Rényi relative entropy implies existence of a recovery map, J. Phys. A: Math. Theor. \textbf{50}, 085303 (2017).

\bibitem{Jencova2018}
A. Jenčová, Rényi relative entropies and noncommutative $L_p$-spaces, Ann. Henri Poincaré \textbf{19}, 2513 (2018).

\bibitem{Jencova2021}
A. Jenčová, Rényi relative entropies and noncommutative $L_p$-spaces II, Ann. Henri Poincaré \textbf{22}, 3235 (2021).

\bibitem{Petz1996}
D. Petz, Monotone metrics on matrix spaces, Linear Algebra Appl. \textbf{244}, 81 (1996).

\bibitem{Robin2008}
R. Blume-Kohout, H. K. Ng, D. Poulin, and L. Viola, Characterizing the structure of preserved information in quantum processes, Phys. Rev. Lett. \textbf{100}, 030501 (2008).

\bibitem{Robin2010}
R. Blume-Kohout, H. K. Ng, D. Poulin, and L. Viola, Information-preserving structures: A general framework for quantum zero-error information, Phys. Rev. A \textbf{82}, 062306 (2010).

\bibitem{Jencova2024}
A. Jenčová, Recoverability of quantum channels via hypothesis testing, Lett. Math. Phys. \textbf{114}, 31  (2024).

\bibitem{Audenaert2008}
K. M. R. Audenaert, M. Nussbaum, A. Szkoła, and F. Verstraete, Asymptotic error rates in quantum hypothesis testing, Commun. Math. Phys. \textbf{279}, 251 (2008).

\bibitem{BarnumKnill2002} 
H. Barnum and E. Knill, Reversing quantum dynamics with near-optimal quantum and classical fidelity, J. Math. Phys. \textbf{43}, 2097 (2002).

\bibitem{Wilde2015}
M. M. Wilde, Recoverability in quantum information theory, Proc. R. Soc. A \textbf{471}, 20150338 (2015).

\bibitem{DupuisWilde2016}
F. Dupuis and M. M. Wilde, Swiveled Rényi entropies, Quantum Inf. Process. \textbf{15}, 1309 (2016).

\bibitem{JungeWilde2016}
M. Junge, R. Renato, D. Sutter, M. M. Wilde, and A. Winter, Universal recoverability in quantum information, in \textit{2016 IEEE International Symposium on Information Theory (ISIT)} (IEEE, 2016), pp. 2494-2498.

\bibitem{Liulizhuo2025}
L. Liu and C. C. Aw, Quantifying irreversibility via Bayesian subjectivity for classical and quantum linear maps, Phys. Rev. E \textbf{112}, 054123 (2025).

\bibitem{LuoSun2024}
S. Luo and Y. Sun, Quantifying the irreversibility of channels, Theor. Math. Phys. \textbf{218}, 426 (2024).

\bibitem{Liyuan2025}
Y. Li, S. Luo, Y. Sun, and S. Wang, Quantifying information loss of quantum channels, Phys. Rev. A \textbf{112}, 062404 (2025).

\bibitem{LiYuan2025a}
Y. Li, N. Liu, and S. Wang, Reversible channels and isometric properties of quantum channels, J. Math. Phys. \textbf{66}, 053505 (2025).

\bibitem{LiYuan2025b}
Y. Li, S. Luo, and Y. Sun, Structure of reversible quantum channels, Ann. Phys. (Berlin) \textbf{538}, e00473 (2026).

\bibitem{LiYuan2026}
X. Sun, S. Gao, and Y. Li, Isometric properties of completely positive maps, Positivity \textbf{30}, 42 (2026).

\bibitem{Holevo2012}
A. S. Holevo, \textit{Quantum Systems, Channels, Information: A Mathematical Introduction} (De Gruyter, Berlin, 2013).

\bibitem{JFL2013}
M. Jiang, S. Luo, and S. Fu, Channel-state duality, Phys. Rev. A \textbf{87}, 022310 (2013).

\bibitem{compatibility2017}
T. Heinosaari and T. Miyadera, Incompatibility of quantum channels, J. Phys. A: Math. Theor. \textbf{50}, 135302 (2017).

\bibitem{GuoYi2025}
Y. Guo and S. Luo, Irreversibility versus incompatibility of quantum channels, Commun. Theor. Phys. \textbf{78}, 045102 (2025).

\bibitem{Choi2009}
M.-D. Choi, N. Johnston, and D. W. Kribs, The multiplicative domain in quantum error correction, J. Phys. A: Math. Theor. \textbf{42}, 245303 (2009).

\bibitem{Ticozzi2010}
F. Ticozzi and L. Viola, Quantum information encoding, protection, and correction from trace-norm isometries, Phys. Rev. A \textbf{81}, 032313 (2010).

\bibitem{Busch1999}
P. Busch, Stochastic isometries in quantum mechanics, Math. Phys. Anal. Geom. \textbf{2}, 83 (1999).

\bibitem{Molnar2002}
L. Molnár and W. Timmermann, Isometries of quantum states, J. Phys. A: Math. Gen. \textbf{36}, 267 (2002).

\bibitem{Stinespring1975}
W. F. Stinespring, Positive functions on C*-algebras, Proc. Am. Math. Soc. \textbf{6}, 211 (1955).

\bibitem{Girard2021}
M. Girard, M. Plávala, and J. Sikora, Jordan products of quantum channels and their compatibility, Nat. Commun. \textbf{12}, 2129 (2021).

\bibitem{GL2}
Y. Guo and S. Luo, Temporal correlating power of quantum channels, Phys. Rev. A \textbf{111}, 062415 (2025).

\bibitem{Umegaki1962}
H. Umegaki, Conditional expectation in an operator algebra. IV. Entropy and information, Kodai Math. Sem. Rep. \textbf{14}, 59 (1962).

\bibitem{vedral2002relative}
V. Vedral, The role of relative entropy in quantum information theory, Rev. Mod. Phys. \textbf{74}, 197 (2002).

\bibitem{Muller2013}
M. Müller-Lennert, F. Dupuis, O. Szehr, S. Fehr, and M. Tomamichel, On quantum Rényi entropies: A new generalization and some properties, J. Math. Phys. \textbf{54}, 122203 (2013).

\bibitem{Wilde2013}
M. M. Wilde, A. Winter, and D. Yang, Strong converse for the classical capacity of entanglement-breaking and Hadamard channels via a sandwiched Rényi relative entropy, Commun. Math. Phys. \textbf{331}, 593 (2014).

\bibitem{luo2011decoherence}
S. Luo and N. Li, Decoherence and measurement-induced correlations, Phys. Rev. A \textbf{84}, 052309 (2011).

\bibitem{Kuperberg2003}
G. Kuperberg, The capacity of hybrid quantum memory, IEEE Trans. Inf. Theory \textbf{49}, 1465 (2003).


%
%
%
%
%
%
%
%
%
%
%
%

\bibitem{ArakiLieb1970}
H. Araki and E. H. Lieb, Entropy inequalities, Commun. Math. Phys. \textbf{18}, 160 (1970).

\bibitem{CarlenLieb2012}
E. A. Carlen and E. H. Lieb, Bounds for entanglement via an extension of strong subadditivity of entropy, Lett. Math. Phys. \textbf{101}, 1 (2012).

\bibitem{JiangLiang2024}
G. Zheng, W. He, G. Lee, and L. Jiang, Near-optimal performance of quantum error correction codes, Phys. Rev. Lett. \textbf{132}, 250602 (2024).

\end{thebibliography}
\end{document}